\documentclass[letter,11pt]{article}

\usepackage[pdftex]{graphicx}

\usepackage{geometry}
\usepackage{natbib}
\usepackage[doublespacing]{setspace}
\usepackage[page]{appendix}

\usepackage{subfig}
\usepackage[clockwise]{rotating}
\usepackage{portland}
\usepackage{float}
\usepackage{amsmath}
\usepackage{verbatim}
\usepackage{mathpazo}
\usepackage{caption}

\providecommand{\U}[1]{\protect\rule{.1in}{.1in}}
\DeclareMathOperator\logit{logit}
\DeclareMathOperator\Prob{Prob}

\bibpunct[, ]{(}{)}{;}{a}{}{,}
\begin{document}

\title{Scalable Inference of Customer Similarities from Interactions Data using Dirichlet Processes}
\author{Michael Braun \\
MIT Sloan School of Management\\
Massachusetts Institute of Technology\\
Cambridge, MA 02139 \\
braunm@mit.edu
\and
Andr\'e Bonfrer \\
College of Business and Economics \\
Australian National University\\
Canberra ACT 0200\\
andre.bonfrer@anu.edu.au
\thanks{The authors thank David Dahl, Daria Dzyabura , Pete Fader,  Jacob Goldenberg, John Hauser, Barak Libai,  Jon McAuliffe, Carl Mela, Adrian Raftery, David Schweidel, and Romain Thibaux for useful suggestions and helpful comments on previous versions of this paper, as well as Rico Bumbaca and Alex Riegler for research assistance, and Jeongwen Chiang and China Mobile for providing the dataset.}}
\date{December 21,  2010}
\begin{singlespace}
  \maketitle
\begin{center}
\vspace{0.6 cm}
\vspace{1.2 cm}
\end{center}
\begin{abstract}
Under the sociological theory of homophily, people who are similar to one another are more likely to interact with one another.
Marketers often have access to data on interactions among customers from which, with homophily as a guiding principle,
inferences could be made about the underlying similarities. However, larger networks face a quadratic
explosion in the number of potential interactions that need to be modeled.  This scalability problem renders probability models of social interactions
computationally infeasible for all but the smallest networks.  In this paper we develop a probabilistic framework for modeling
customer interactions that is both grounded in the theory of homophily, and is flexible enough to account for random variation in who interacts with whom.  In particular, we present a novel Bayesian nonparametric approach, using Dirichlet processes, to moderate the scalability problems that marketing researchers encounter when working with networked data.  We find that this framework is a powerful way to draw insights into latent similarities of customers, and we discuss how marketers can apply these insights to segmentation and targeting activities.   
\end{abstract}
\end{singlespace}
\thispagestyle{empty}
\newpage

\section*{Introduction}
Marketers have long been interested in the notion that interactions among customers will affect behavior.  For example, knowledge of how customers relate to one another improves our understanding on how preferences are formed \citep{ReingenFosterEtAl1984}, how preferences are correlated within groups \citep{WittBruce1972, ParkLessig1977, FordEllis1980, BeardenEtzel1982}, or how useful referrals are for marketers when developing new markets \citep{ReingenKernan1986}.  Connections among customers are opportunities for preference influence (e.g. contagion in diffusion, \citealt{Bass1969}).  Marketers can leverage "word of mouth" to amplify the efficacy of their communication campaigns \citep{GoldenbergLibaiMuller2001,NamManchandaChintagunta2007, IyengarVandenBulteValente2008, GodesMayzlin2009}.  Incorporating network information into marketing models has also been shown to improve forecasts of both new product adoption \citep{HillProvost2006} and customer churn \citep{DasguptaSingh2008}.   

Similar customers are more likely to interact with one another, so given the need for marketers to find efficient ways to attract and cultivate customers, there exists vast opportunity in leveraging interactions data to infer similarity and connect this to marketing behavior (e.g., \citealp{YangAllenby2003,BellSong2007,NamManchandaChintagunta2007}).  This link between similarity and interactions is the sociological theory of homophily \citep{Akerlof1997,Blau1977,LazarsfeldMerton1954} 
and is the basis for many marketing studies that examine or accommodate interactions among customers (e.g. \citealp{GatignonRobertson1985,BrownReingen1987,ChoiHuiBell2010}).  Put simply, homophily implies that customers who are similar to one another are more likely to interact with one another, and share information and influence, than customers who are not.  There is a substantial volume of literature that links similarities to interactions (see \citealt{McPhersonSmithLovin2001}, for a review), but interactions and similarities are not the same thing.  We consider interactions to be the ``data'' that records some observable action between two individuals, while similarities form a latent, unobserved construct (though possibly correlated with other observed measurements) that determines which individuals are more likely to interact with others. 
In this paper, we present an illustrative yet parsimonious model, grounded in the theory of homophily, that allows marketers to infer  latent similarities from observed interactions.  The idea is to develop a probability model that uses interactions data to infer latent similarities, and generates output that can help marketers better understand why customers interact with whom they do, or why they behave the way they do, in terms that are useful to marketers.  

We build on a class of probability models known as latent space models \citep{HoffRafteryHandcock2002,HandcockRaftery2007}.   The fundamental idea behind latent space models is that each individual is characterized as occupying some unobserved point on a multi-dimensional space.  When estimated on relationship data (e.g. a list of self-reported friendships, as in \citealt{ReingenFosterEtAl1984,BrownReingen1987}, or working relationships, as in \citealt{IyengarVandenBulteValente2008}), the distance among points in latent space determines the probabilities for the incidence of these relationships.\footnote{Latent space methods are, of course, not limited to examining social network data, and could be used to model similarities between units in two distinct groups \citep{BradlowSchmittlein2000}, or to model the difference in knowledge by individuals \citep{VanAlstyneBrynjolfsson2005}.  Further applications are discussed in \citet{ToivonenKovanen2009}.}  What is becoming increasingly available to marketers, however, are clean, observational data on the interactions among customers, such as phone call records or online social networking transactions, but with no observed information about the content of the interaction (who these people are and what they talk about), or the nature of the relationship between these individuals (what it is about these two particular people that generates an interaction between them). 

When latent space models are estimated on interactions data, we can interpret the distance among points as relative similarity.   Homophily gives us the theoretical foundation on which we can make this claim.  The managerial usefulness of estimating latent space models on interactions data comes from identifying and inferring these similarities.  Sometimes, such as our application in telecommunication services, interactions generate revenue directly.  There are many examples, like those mentioned in the first paragraph, where marketers deliberately target customers who will contact, and hopefully influence, others.  But in other cases, the marketing activities themselves might have nothing at all to do with ``following network links,'' or generating ``word of mouth.''  Knowing how similar customers are to one another is of direct relevance to marketing practitioners because it forms the basis of segmentation and targeting across a heterogeneous population.  Once we have inferences about relative similarities of customers in hand (through posterior distributions of latent distances), we can segment and target customers accordingly.  Ordinarily, this segmentation is done based on observed characteristics of individuals.  Very little attention has been paid to how marketers might be able exploit the information contained in interactions data for traditional, non-networked marketing tactics, like deciding in which publications (online or otherwise) to advertise.  Indeed, the company that uses interactions data for segmentation and targeting (e.g. an online retailer) does not necessarily have to be the same company that collects it (e.g., the cell phone provider).

One reason modelers have not been able to apply latent space models to marketing data in a general sense is that it can be a daunting computational challenge.  One of the key tenets of probability modeling is that we need to take all data into account, including pairs of individuals for whom we do not observe any interactions at all (the ``zeros'' in the data offer valuable information about relative similarities).  Thus, there has been a formidable obstacle to using probability models for larger observational network datasets.  A dataset with $N$ individuals involves $\binom{N}{2}$ dyads (the binomial coefficient $\binom{N}{x}$ is defined as $\frac{N!}{x!(N-x)!}$).  For the exemplar dataset that we use in this paper, there are 11,426,590 sets of dyad-specific parameters that we need to consider, and this is for a dataset of only 4,781 individuals.  Unless we want to break the interdependencies among dyads, ignore unobserved heterogeneity, or make other assumptions that are similarly restrictive, we need to compute all of these $\binom{N}{2}$ dyad-specific likelihoods, and the same number of dyad-specific parameters, at each iteration of our estimation algorithm.  The problem with scale makes probability models of social interactions computationally intractable for all but the smallest datasets.

The modeling challenge is therefore to reveal similarities in heterogeneous characteristics from customers' interaction data, in a scalable and interpretable way.  We accomplish this by applying a Bayesian nonparametric prior, the Dirichlet process (DP), as the distribution of locations on the latent space.  The DP is essentially a distribution over distributions (as opposed to over scalars or vectors), and for our purposes, its most salient characteristic is that each realized distribution is \emph{discrete}.  Consequently, individuals in the network are clustered on common locations on the latent space.  So if this discrete distribution has $k$ mass points, there are only $\binom{k}{2}+1$ distinct distances on the latent space (the $+1$ comes from the zero distance between two individuals at the same latent coordinate).  Since $k$ must be smaller than $N$,  there are substantially fewer distinct likelihoods to compute and parameters to estimate.

In this research, we show how marketers could use latent space models to segment customers based on posterior inferences of latent similarities, using this more efficient Bayesian nonparametric approach.  An output of our algorithm is a posterior estimate of the latent space that is inferred from the interactions data.   Our probabilistic approach to modeling these data allows for the fact that similar individuals may not interact, even though they may have similar characteristics and travel in the same social circles.  Also, we recognize that while interactions typically occur among similar customers, there is also the possibility that dissimilar customers (who may have different purchase patterns and preferences) may interact at some time.  To demonstrate the power and utility of this approach to modeling interactions data, we apply it to a dataset of observed interactions from a cellular communication network.   We propose a probability specification for this particular dataset, in which the incidence and rates of interactions are functions of distances in latent space.  We validate the approach in two ways:  by showing that adding the latent space structure to the probability model improves the fit of the model, with respect to several metrics commonly used in the social networking literature; and by showing that the latent space model can distinguish among pairs of individuals for whom the observed number of interactions are all \emph{identically} zero during a calibration period, in terms of how well the model predicts which of those pairs will eventually interact in a future holdout period.  These tests demonstrate that failing to account for the unobserved heterogeneous interdependencies among individuals leads to a model that simply does not represent the observed patterns in interactions.

We then assess the computational improvements and scalability issues surrounding our Bayesian nonparametric approach, and the managerial insights that one can get from estimates of the latent space itself.  By using a graphical representation of the latent space, we show how marketers can augment network based practices that follow observed interaction paths, with tactics that segment and target customers according to inferred latent similarities.  The data that are available to us do not let us offer hard evidence of a correlation between similarities and purchase preferences but, given the findings in the marketing literature that show the importance of similarities and interactions in customer behavior, it is reasonable to expect that marketing mix efforts benefit from being able to distinguish interactions among similar customers from interactions among dissimilar customers.  The computational improvements from using a DP prior for the latent space make these inferences attainable for the datasets that marketers typically encounter.

\section{General model formulation\label{section:model}}

\subsection{Intuitive description}\label{subsection:modelIntuition}

In a probability model of network data, each dyad in the network generates some vector of data, which can represent a wide variety of behavior.  Examples include binary indicators of relationships, counts of transactions (among customers), times between interactions, or combinations thereof.  However simple or complicated the data are, they should be treated as some output of a stochastic process that is governed by dyad-specific parameters (and possibly some additional population-level parameters).  Data that is generated from a network of customers differs from individually-generated data, such as household purchase data, in that we can no longer assume that the data-generating processes are independent across dyads.   For example, if we were to observe telephone calls between members of a dyad, the rate at which A calls B, and B calls C, can provide information about how often A calls C.  However, we do assume that the dyad-level processes are \emph{conditionally} independent, so the only correlation among dyads is what occurs because of similarities in parameters.   This means that even though frequencies of phone calls might be dependent across dyads, the specific times at which those calls ultimately take place are independent, conditional on the rate of interactions.

We determine dyad-level parameters so that similar individuals will have higher incidence of interaction than dissimilar individuals.   The characteristics upon which this similarity is based are likely unobservable by the researcher.  Therefore, we represent unobserved, exogenous characteristics of the individual (and thus, the individual himself), as a $D$-dimensional vector on some latent space 
\citep{HoffRafteryHandcock2002,HandcockRaftery2007,BradlowSchmittlein2000,VanAlstyneBrynjolfsson2005}.   Similarity between 
two individuals is 
measured by the distance between their latent coordinates across this latent space, and we can express the rates 
or probabilities of interaction between two people as a decreasing function of the latent distance between them.  
Note that these distances and locations do not directly represent physical or geographic locations in any way (although they may, of course, be incidentally correlated with them).  Instead, they are individual-level
parameters to be estimated, based on observed patterns of interaction.  For the purposes of 
this article, we treat the location of each latent coordinate as persistent and stationary.  So even though interactions among people 
may appear and disappear periodically (a non-stationary observed phenomenon that \citet{KossinetsWatts2006} describe as \emph{evolving}), the underlying rates and probabilities of these incidences remain the same.  Thus, our stationary model can still capture non-stationary behavior in the observed data.
Also, we want to emphasize that a latent space model is an abstraction of reality, and we caution researchers 
not to place too much concrete meaning on any one dimension.  
It is the relative distances among individual latent coordinates, and not the absolute positioning in the latent space, that matter.

\subsection{Formal model}\label{subsection:modelFormal}

A more formal definition of the general model is as follows.  Let $y_{ij}$ be a vector of observed data that is attributable to the dyad of person $i$ and person $j$, and let $f\left( y_{ij} |\theta_{ij}\right)$ be the likelihood of observing $y_{ij}$, given the dyad-specific parameter vector $\theta_{ij}$.  Next, let $\theta_{ij}$ be heterogeneous across dyads, with each $\theta_{ij}$ drawn randomly from a dyad-specific prior distribution $g\left(\theta_{ij} | \phi_{ij}\right)$.  A model in which $\phi_{ij}$ is common across all dyads, or itself distributed independently (drawn from its own mixing distribution), would imply cross-dyad independence of  $\theta_{ij}$, which may not make sense in a network setting.  To incorporate some network-based dependence in the distribution of $\theta_{ij}$, we instill a pattern of heterogeneity of $\phi_{ij}$ that allows for a useful, intuitive interpretation of the similarities.  Thus, there are two sources of heterogeneity that generate $\theta_{ij}$:  independent dyad-level variation from $g\left( \theta_{ij} | \phi_{ij}\right)$, and network-induced interdependence in the distribution of $\phi_{ij}$. 

Before explaining how we model heterogeneity in $\phi_{ij}$, let us shift our focus from the level of the dyad to the level of the individual.  Each dyad is made up of two individuals, each of whom has its own, mostly unobserved traits and characteristics.  Let $z_i$ be a $D-$dimensional vector that is associated with person $i$, and let $z$ be the collection of all $N$ of these vectors.  Since each $z_i$ is unobserved, we call it a ``latent coordinate,'' and the $D$-dimensional space on which it lies a ``latent space,'' as in \citet{HoffRafteryHandcock2002} and \citet{HandcockRaftery2007}.  Even if the $N$ vectors in $z$ are distributed independently on the latent space, the \emph{distances} between every \emph{pair} of $z_i$  (the ``latent distances'') are not.  By expressing $\phi_{ij}$ as a monotonic function of the distance between $z_i$ and $z_j$, we induce dependency among all the $\phi_{ij}$ and, in turn, all the $\theta_{ij}$.  As an example, suppose that $\theta_{ij}$ represents a rate of contact between $i$ and $j$, and the distribution of $\theta_{ij}$ depends positively on $\phi_{ij}$ (for example, the mean of  $g\left(\theta_{ij} | \phi_{ij}\right)$ increases with $\phi_{ij}$).   We determine $\phi_{ij}$ by evaluating a monotonically decreasing function of the latent distance, so as $i$ and $j$ are less similar (the distance between $z_i$ and $z_j$ goes up), the rate of interaction between $i$ and $j$ goes down.  But we never need to estimate $\theta_{ij}$ or $\phi_{ij}$ directly.  We need only to estimate the locations of $z_i$ for all $N$ people to get the values of $\phi_{ij}$ for all $\binom{N}{2}$ dyads.

\subsection{Mixtures of Dirichlet processes:  what they are, and how to use them to model the latent space}\label{subsection:semiparametric}
Even if we model $\phi_{ij}$ as a function of the latent distance among the $z_i$'s, we still have the issue that there are many $z_i$'s, and thus a large number of latent distances, to model.  This means that at each iteration of our estimation algorithm, we need to compute $\binom{N}{2}$ values of $\phi_{ij}$, and $\binom{N}{2}$ corresponding data likelihoods.  When $N$ is small, scalability becomes less of a problem, and one could use the original parametric formulation of the latent space model.  But as $N$ becomes even moderately large, estimating the latent coordinates becomes computationally infeasible.  We reduce the number of distinct values of $z_i$ by using a discrete distribution, $H\left( z_i|\cdot\right)$  for the distribution of $z_i$ on the latent space.  If this discrete distribution has $k$ mass points, then there are only $\binom{k}{2}+1$ distinct latent distances.  For a given network size, a larger difference between $k$ and $N$ leads to a greater computational savings by having fewer distinct values 
of $\phi_{ij}$ to consider. 
 To avoid having to estimate each $\theta_{ij}$ directly, we choose $f\left( y_{ij}|\theta_{ij}\right)$ and $g\left( \theta_{ij} |\phi_{ij}\right)$ such that we can integrate over $\theta_{ij}$ analytically, and express the marginal distribution $f\left( y_{ij} |\phi_{ij}\right)$  in closed form.   However, we do not want to prespecify the functional form of $H$, because we don't know for certain what it is, nor do we want to prespecify $k$, since we do not know what the ``correct'' number of mass points for $H$ is. 

Our approach is to use a mixture of Dirichlet processes as a Bayesian nonparametric prior distribution for the points on the latent space.  Although the properties of Dirichlet processes (DP) have been known for a while (back to \citealt{Ferguson1973}), they are still relatively new to marketing.   The few examples include \citet{AnsariMela2003} (as a Bayesian alternative to collaborative filtering), \citet{KimMenzefrickeFeinberg2004} (identifying clusters of customers in discrete choice models), \citet{WedelZhang2004} (analyzing brand competition across subcategories) and \citet{BraunFaderEtAl2006} (estimating thresholds of claiming behavior for home owners' insurance).   In our context, a Dirichlet process is a probability distribution over distributions (as opposed to a distribution over a scalar or vector).  Accordingly, a single draw from a Dirichlet process is itself a random distribution, from which we can draw samples of a variable of interest.  An important feature for our context is that each realization of a DP is a \emph{discrete} distribution, with its support having a finite number of mass points ($k$ in the previous paragraph), so the DP can be thought of as a prior distribution on discrete distributions.\footnote{\baselineskip 12pt The formal definition of what makes a stochastic distribution a DP is a more technical issue.  Essentially, the probabilities of certain events occurring must follow a Dirichlet distribution with parameters that depend on $H_0$ and $\alpha$).   There is an accessible and readable explanation in \citet[][ch. 13]{OHaganForster2004}. }  

There are, of course, many ways to model discrete points on a space; a traditional latent class model with a prespecified number of locations is an extreme example.  What makes the DP more useful in this context is that it has a parsimonious representation, with straightforward sampling properties, and does not require a prespecification of the number of mass points.  In our latent space framework, we let $H$ be a realization from $DP(H_0,\alpha)$, and then have each $z_i$ be a draw from $H$.  The first parameter,  $H_0$,  is itself a probability distribution, and is the ``mean'' of the distributions that the DP generates.    A scalar $\alpha$ controls the variance of the realizations of the DP around $H_0$.  This variance is low when $\alpha$ is high, so for high $\alpha$, realizations from $DP(H_0,\alpha)$ will look a lot like the distribution function of $H_0$.  This concentration of the DP towards $H_0$ results from a DP that generates a discrete distribution with a lot of mass points (a high $k$).  When $\alpha$ is low, realizations from $DP(H_0,\alpha)$ look much less like $H_0$ (high variance), because this DP generates discrete distributions with fewer mass points.  So $\alpha$ plays an important role in determining just how discrete (i.e. value of $k$), or clustered, a DP-generated distribution really is.  Reasonable choices for $H_0$ are those distributions for $z_i$ that one might use in a purely parametric model (note that $H_0$ could have parameters of its own, with their own priors, that need to be estimated.)  Depending on the application, one can either put a prior on $\alpha$ or set it directly.

Given $H_0$ and $\alpha$ we need to know how to simulate $H$ from $DP(H_0,\alpha)$ and then each $z_i$ from $H$.  Since $H$ is nonparametric, even though we know it was generated by the $DP(H_0,\alpha)$, the posterior distribution of any new $z_i$ depends on all the other $z_{-i}$.  Consequently, there is no obvious way to draw a $z_i$ from $H$ directly.  The ``trick'' is to integrate out $H$ analytically, and treat $z_i$ as if it were drawn from this marginal distribution, a mixture of Dirichlet processes (MDP, see \citealt{Antoniak1974}).   The probability of any one $z_i$, given the empirical distribution $ED(\cdot)$ of all the other $z_{-i}$, is
\begin{align}
  \label{eq:MDPprior}
  \Prob(z_i|z_{-i}, H_0, \alpha)&=\frac{\alpha H_0 + ED(z_{-i})}{\alpha+N-1}
\end{align}\citep{BlackwellMacQueen1973,Escobar1994}. So in the estimation algorithm, $\alpha$ determines how likely it is that any new draw of $z_i$ comes from one of the existing, distinct values already possessed by another individual in the dataset (if this is likely, then there are few mass points, with lots of clustering), or from the baseline distribution $H_0$, as a new value.   

\begin{figure}
  \centering
 \subfloat[Sample realizations]{
   \includegraphics[scale=0.52]
   {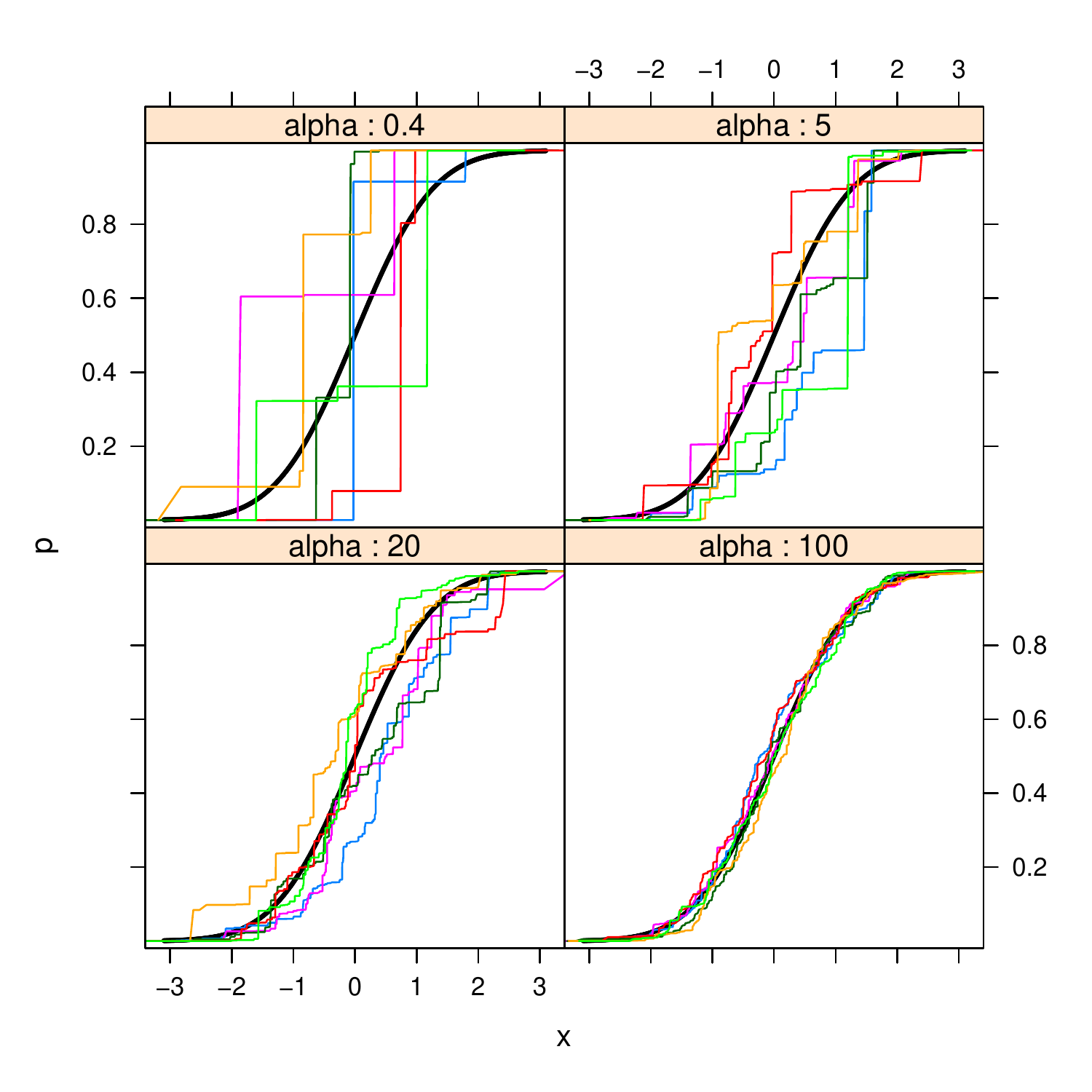}\label{fig:sampDP-cdf}
  }
 \subfloat[Histograms of draws from a single realization]{
    \includegraphics[scale=0.5]
    {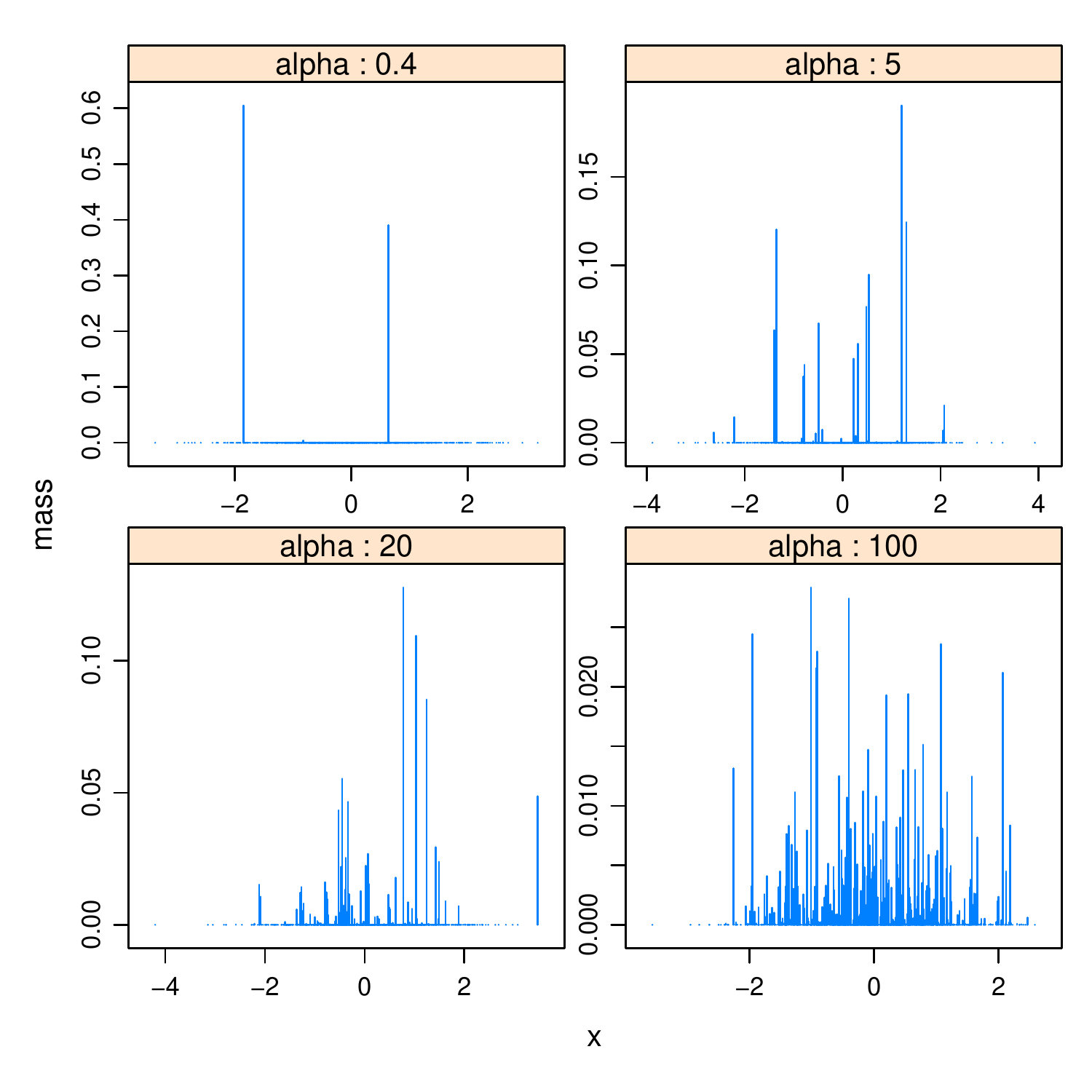}\label{fig:sampDP-mass}
  }\\
\caption{Illustration of realizations from a mixture of Dirichlet process with $H_0$ being a standard normal.  At left, the black line is the cdf of the $H_0$, and each colored line represents a single realization of a MDP.  Each panel corresponds to a different value of $\alpha$.  At right, each panel is a histogram of draws from a single realization.}
\label{figure:DP1}
\end{figure}

To illustrate how this works, Figure \ref{figure:DP1} shows simulations from an MDP when $H_0$ is a univariate standard normal distribution, for different values of $\alpha$.  In the figure, the heavy black line is the standard normal cdf, and each colored line is a single realization from the MDP.  We see that when $\alpha$ is low, there are fewer mass points in each realization, and when $\alpha$ is high, the higher number of mass points allows the realizations to approximate the normal cdf.  In Figure \ref{fig:sampDP-mass}, for each $\alpha$ we present histograms from draws of a single realization of the MDP (so these are draws from a distribution that the MDP generated).  Again, we see fewer distinct clusters (low $k$) when $\alpha$ is low, more clusters when $\alpha$ is high.  In our network model, we are dealing with more dimensions and a richer specification of $H_0$, but the basic idea remains the same.

How we select $H_0$ and $\alpha$, and the priors we place on them, is described in more detail in Appendix \ref{app:posterior}.  Selection of an appropriate distribution for $H_{0}$ requires that we introduce some identifying restrictions on the location vectors ($z_i$).   The concern is that we cannot simultaneously and uniquely identify both the scale of the latent space, and the parameters of the distance function determining $\phi_{ij}$.   To handle this problem, we constrain the prior distribution of $z_i$ so the mean
distance of any $z_{i}$ from the origin is one.  However, we need to
do this without introducing too much ``incorrect''
prior information.  For example, a simple choice for $H_0$ could be a standard multivariate normal
distribution; setting the mean at the origin and the variance at one addresses the translation
and scale identification issues.   The problem with defining
$H_{0}$ as a multivariate normal is that it implies that our prior on the
\emph{\normalsize distribution} of $z_i$ has a mode at the origin. This prior
turns out to be informative, as it generates artifactual clusters of
individuals around the origin in the posterior.  An alternative
specification for $H_{0}$ could be a bounded uniform distribution (so
the mean distance from the origin remains one), but that would
constrain all $z_{i}$ to be inside a hypersphere, effectively placing an
upper bound on the latent distance between customers. This, too, seems like
an unreasonable expression of prior information.  Our solution 
involves using spherical coordinates for $z_i$, consisting of two components: a radius representing the distance from the origin,
and the location on the surface of a hypersphere that has that radius.  We show in Appendix \ref{app:posterior} that $H_0$ can be factored into priors for these two components from which it is straightforward to draw samples.    

This prior on $z_i$, combined with the data likelihood, leads to conditional posterior distributions that are easily incorporated into Gibbs samplers.   \citet{Escobar1994} and \citet{EscobarWest1998} describe some of the theory and derivations behind this, while \citet{Neal2000} details step-by-step instructions on how to add MDPs to Gibbs samplers for both conjugate and nonconjugate models.\footnote{This way of expressing the MDP (and the approach we took in our estimation  is known as the ``Polya urn'' representation.  There is another, equally useful approach known as the ``stick-breaking'' representation \citep{Sethuraman94} that one can also use to build conditional posterior distributions for Gibbs samplers \citet{IshwaranJames2001}.}  
Thus, MDPs allow marketing modelers to relax many of their distributional assumptions by adding only one additional step to the parametric Gibbs sampling algorithm.   We give details of our estimation algorithm in Appendix \ref{app:SamplerAppendix}.  Our exploitation of the discreteness property of Dirichlet processes also lets us reduce the computational burden substantially, as we demonstrate in Section \ref{section:scalability}.  

\section{Example:  telephone calls}\label{section:phonecalls}
We now turn to a specific application of our model, using a dataset
provided by Chongqing Mobile, a subsidiary of China Mobile, 
the largest cellular phone operator in China.  Cellular phone networks have been reported to be highly representative
of self-reported friendships \citep{EaglePentlandLazer2009}, making such data ideal 
for studies of network based interdependencies among customers.
The
data consist of contact record information (for phone calls and SMS messages)
for a panel of $4,781$ residents of Chongqing who are members of the ``silver
tier'', ``gold tier'' or ``diamond tier'' of
the company's preferred customer program.  Each record contains the identifiers for both parties in the contact,
and the date of when the contact takes place. For the purposes of this example, we ignore contacts with people outside this
$N$-person network.\footnote{Our intent in using this dataset is to demonstrate the 
effectiveness of our estimation method, and to illustrate some of the issues that arise 
when modeling dyadic data.  Therefore, we treat our dataset as an entire population of 
individuals, and not as a random sample; our interest is only in contacts made among 
individuals in this population.  
If we were to generalize parameter estimates and predictions to a greater population, ignoring out-of-network 
calls could influence specific parameter estimates.   There are an additional 209 silver, gold or diamond customers in the panel for whom there were no observed calls to other silver, gold or diamond customers during the observation period.} The observed geodesic distance is finite for all dyads (i..e, all customers are connected to every other customer in a finite
number of steps).   We divide the observation period into
a six-month calibration period and a six-month holdout period.
Descriptive statistics for this dataset are summarized in Table
\ref{table:summaryStats}.  Of the $18,078$ nonempty dyads in the
dataset, only $7,559$ appear in both the calibration and holdout
samples.  $5,058$ dyads are nonempty in calibration, but empty in
holdout, and $5,461$ are empty in calibration, but nonempty in holdout.

\begin{table}\centering

\begin{tabular}{l|rrr}
  &Calibration &Holdout &Full\\
\hline
Weeks &26&26&52\\
Customers &4,781&4,781&4,781\\
Non-empty dyads &12,617&13,020&18,078\\
Proportion of empty dyads &0.9989&0.9989&0.9984\\
Clustering coefficient&0.128&0.127&0.127\\
Mean (s.d.) degree distribution &5.3 (4.9)&5.4 (5.3)&7.6 (6.7)\\
Mean (s.d.) geodesic distance&5.5 (1.4)&5.3 (1.3)&4.7 (1.1)\\
Mean (s.d.) calls per non-empty dyad&7.5 (18.7)&7.6
(18.7)&15.1 (36.0)\\
Mean (s.d.) shared friends in non-empty dyad
   &3.0 (2.7)&3.3 (2.8)&5.7 (2.8)\\
\end{tabular}
\caption{Descriptive statistics of China Mobile dataset.}
\label{table:summaryStats}
\end{table}

One way to describe the structure of the observed network is to compare it to the ``small world'' networks described in \citet{WattsStrogatz1998} and \citet{Watts1999}.  Generally speaking, a small world network is one in which everyone in the network is connected to everyone else through a relatively small number of intermediaries (i.e., a low mean geodesic distance), and a relatively large number of common friends who are connected among themselves (i.e., a high clustering coefficient).  We can assess the extent to which a network is ``small world'' comparing the mean geodesic distances and clustering coefficients to those that we would expect to see from a network in which connections are determined at random for the same number of people (4,781) and average number of ``friends'' per person (7.6).  Using the asymptotic approximations in \citet{Watts1999}, the mean geodesic distance we would expect from a random graph of this size is about 4.2, and the expected clustering coefficient is about 0.002.  In the observed Chongqing Mobile network we observe quite a bit more clustering than we expect to see from a random graph, while the mean geodesic path is slightly longer what we would expect.  One possible reason that our mean geodesic distance is not smaller is that we could have a large number of small clusters, and not all small clusters are connected to each other.   In fact, our estimates of  $k$ (illustrated in Figure \ref{figure:scalePlots}) will bear this out.  We also note that our network would not qualify as a ``scale free'' network, in that the degree distribution clearly does not follow a power law-type distribution (we show the observed degree distribution in Figure  \ref{figure:PPC1}).

\subsection{Model specifics}
\label{subsection:modelSpecifics}
Using the notation introduced in Section \ref{section:model}, $y_{ij}$ is the vector of intercontact times, ending with the survival time (the duration between the last observed contact and the end of the observation period).  If there are no observed contacts in the dyad, $y_{ij}$ is the length of the observation period, and we call that dyad ``empty.''  If there are observed calls, the dyad is ``non-empty.''  The definition of $f\left(y_{ij}|\theta_{ij}\right)$ follows the logic of the ``exponential never-triers'' model in \citep{FaderHardieZeithammer2003}, which in turn draws from the ``hard-core never-buyers'' model in \citet{MorrisonSchmittlein81} and \citet{MorrisonSchmittlein1988}.  First, there is a probability $p_{ij}$ that a dyad will remain forever empty, no matter how long we wait.  We call dyads like this ``closed''.  Next, for dyads that are ``open,'' (with probability $1-p_{ij}$), intercontact times follow an exponential distribution with rate $\lambda_{ij}$.  To link these specifics with our general model, $\theta_{ij}=\{ p_{ij},\lambda_{ij}\}$.  Note that there are two ways we could observe an empty dyad.  The dyad is either closed, or it is open, but with a contact rate that is sufficiently low that we just happened to not observe any contacts during the observation period.

Whether the exponential distribution is appropriate for this dataset is ultimately an empirical question, but we choose it for four reasons.  First, we do not need to make special provisions for left-censoring because of the memorylessness property.  Second, the number of contacts is a sufficient statistic for the individual elements in $y_{ij}$.  We were able to exploit these two features of the exponential distribution to gain computational savings without compromising the fundamental purpose of the research.  Third, we did run the model on a much smaller dataset where $f\left(\cdot\right)$ is governed by a ``Weibull never-triers'' model, in order to allow for duration dependence, and found that since the shape parameter of the Weibull was close to $1$, it reduced to the exponential distribution anyway.  Finally, we chose the exponential distribution because it forms a conjugate pair with our choice of $g\left( \theta_{ij}|\phi_{ij}\right)$, a gamma distribution for $\lambda_{ij}$ with dyad-specific mean $\mu_{ij}$ and common variance $v$, and a degenerate distribution over $p_{ij}$, so that at this level of the hierarchy, $p_{ij}$ is homogeneous for all dyads (we will add heterogeneity to $p_{ij}$ later through the latent space).  The exponential-gamma pair lets us integrate over $\theta_{ij}$ analytically, further easing computational effort.  The vector $\phi_{ij}$ therefore contains three elements, $p_{ij}, \mu_{ij} \text{ and } v$ ($p_{ij}$ is contained in both $\theta_{ij}$ and $\phi_{ij}$.)

To evaluate whether latent space is worth adding to a model of interactions data, we estimated the model with three different definitions of the elements of $\phi_{ij}$.  For a ``Baseline'' model, we let $\phi_{ij}=\phi$, a common value for all dyads (note that we still maintain dyad-level heterogeneity in $\theta$, but it does not appear explicitly in the data likelihood).  For a second model, HMCR (for ``Homogeneous Mean Contact Rate''), $\mu_{ij}$ and $v$ remain homogeneous across dyads, but $p_{ij}$ is now determined by the distribution on the latent space.  Specifically, we define 
\begin{equation}\label{eq:logit_p}
\logit p_{ij} = \beta_{1p} - \beta_{2p} d_{ij}^{\beta_{3p}}
\end{equation}
where the $\beta$'s are coefficients to be estimated, and $d_{ij}$ is the latent distance between $z_i$ and $z_j$.  For a third model, named ``Full,'' $p_{ij}$ retains the same definition as in Equation \ref{eq:logit_p}, except that $\mu_{ij}$ is now heterogeneous across dyads, defined as
\begin{equation}\label{eq:log_mu}
\log \mu_{ij} = \beta_{1\mu} - \beta_{2\mu} d_{ij}^{\beta_{3\mu}}
\end{equation}

Equations \eqref{eq:logit_p} and \eqref{eq:log_mu} allow the respective relationships to latent distance to be concave, linear or convex.  The parameters $\beta_{2p},\beta_{3p},\beta_{2\mu},\text{ and } \beta_{3\mu}$ are constrained to be nonnegative, because as latent distance increases, the probability of contact, and the rate of contact, should decrease.    We selected Euclidean distance as our distance measure, after experimenting with others that did not perform as well \citep{VanAlstyneBrynjolfsson2005}.\footnote{Here, we are talking about distance between two individuals' coordinates on the \emph{latent} space.  This concept of distance is different from when we talk about geodesic distance, which is the smallest number of \emph{observed} connections along the shortest path between two individuals.}  Another candidate for this distance metric is the Mahabalonis distance (as used in \citealt{BradlowSchmittlein2000}), which weights some dimensions more than others in the computation of the distance among individuals.  However, the non-parametric nature of the estimated latent space means the dimensions are already differentially scaled.  Also, the Euclidean distance is computationally more efficient.  As with the parametric specification, the functions in Equations \ref{eq:logit_p} and \ref{eq:log_mu}, and the distance measure, are subject to empirical testing and may not be appropriate in all contexts.

\subsection{Assessing contribution of the latent space}\label{section:ppc}

So far, we have assumed that parameter interdependence is an important characteristic of a model of customer interactions.  However, one could falsify this claim by showing that models in which dyad-level parameters are independent fit no worse than models that incorporate a latent space.  We ran our algorithm with latent spaces of different dimensionality and, based on estimates of log marginal likelihoods, we decided that the parsimonious choice of $D=2$ is most appropriate (see Appendix \ref{app:LML}).   As evidence that the latent space models do better than independent models, we evaluate the contribution of latent space based on both posterior predictive checks (PPCs) and on forecasting interactions in empty dyads. 

\begin{figure}
  \centering
  \subfloat[Degree Distribution]{
   \includegraphics[height=45mm,width=120mm]
    {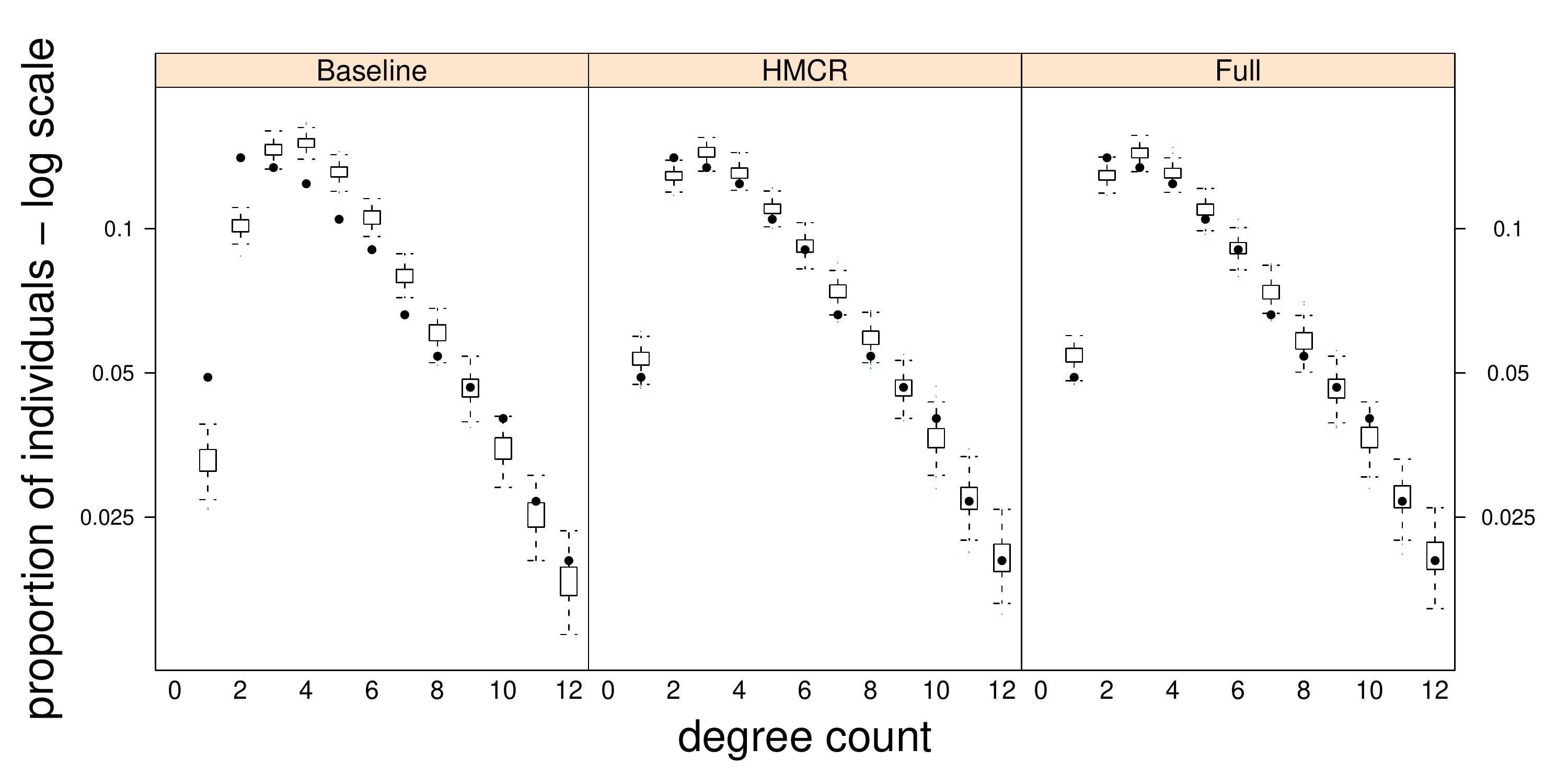}
  }\\\vspace{-0.3in}
  \subfloat[Dyadwise Shared Partners Distribution]{
   \includegraphics[height=45mm,width=120mm]
    {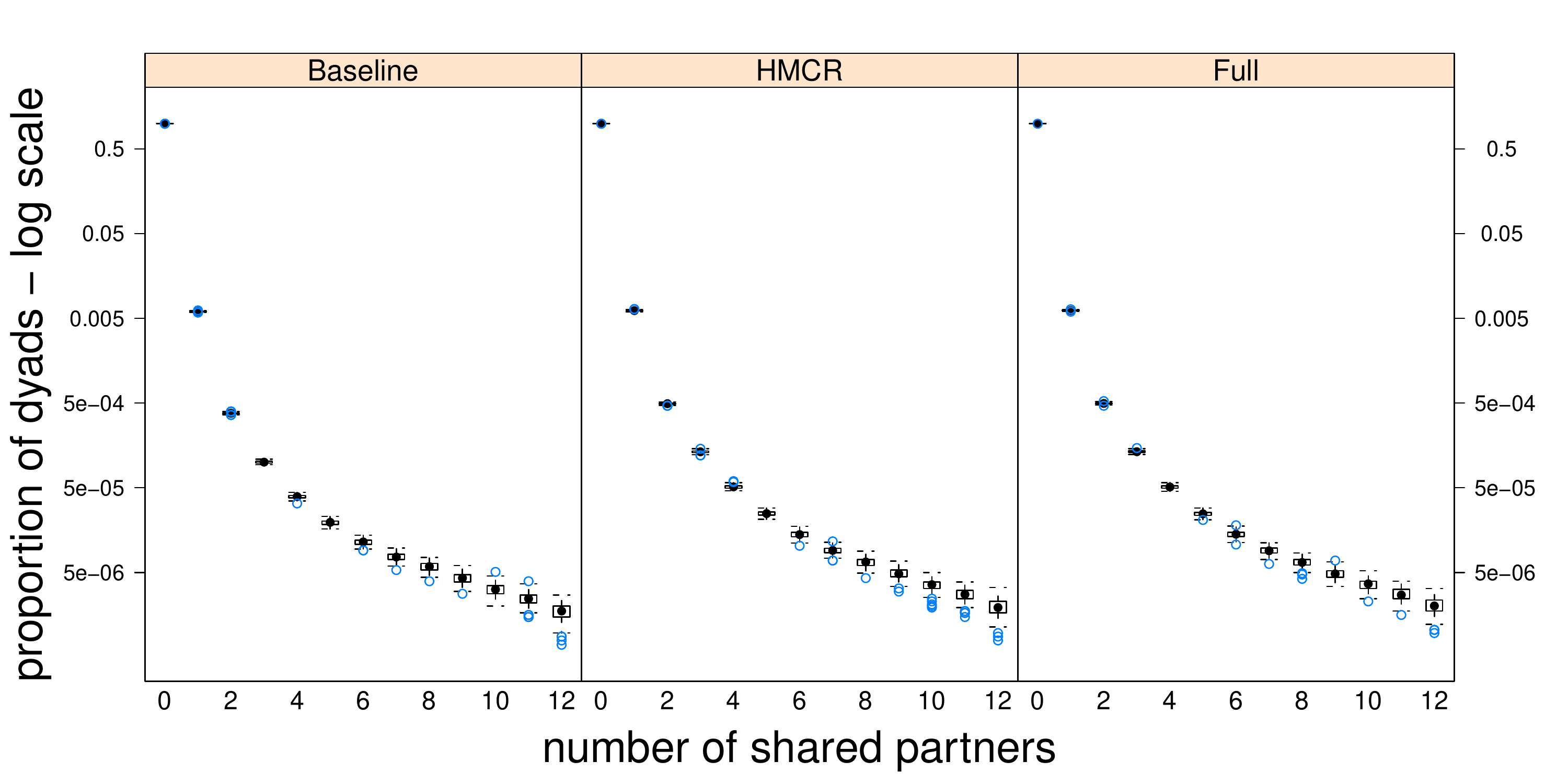}
  }\\\vspace{-0.3in}
  \subfloat[Geodesic distance distribution]{
   \includegraphics[height=45mm,width=120mm]
    {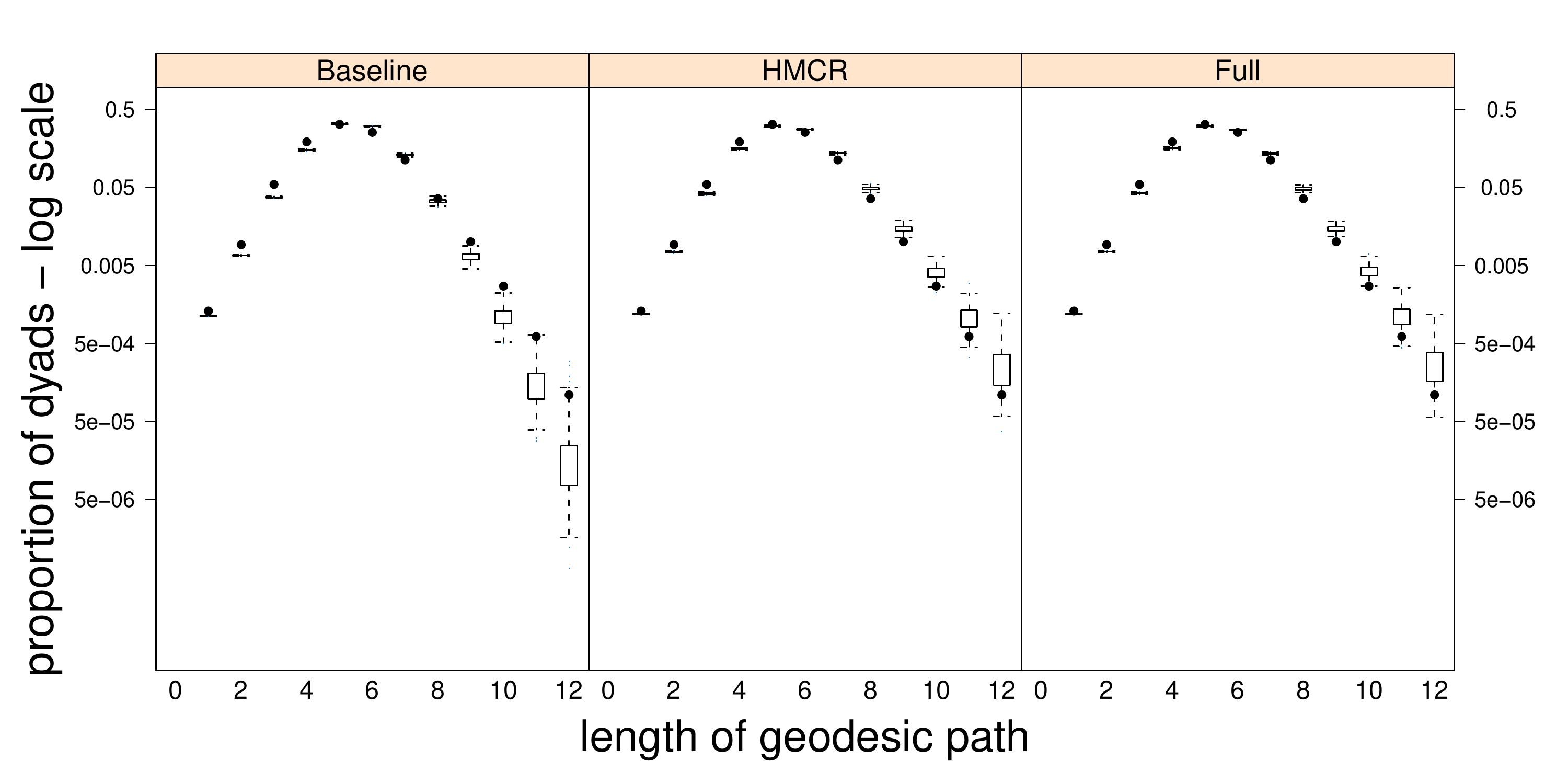}
  }\\\vspace{-0.5in}
\subfloat[Number of calls]{
   \includegraphics[height=60mm,width=120mm]
    {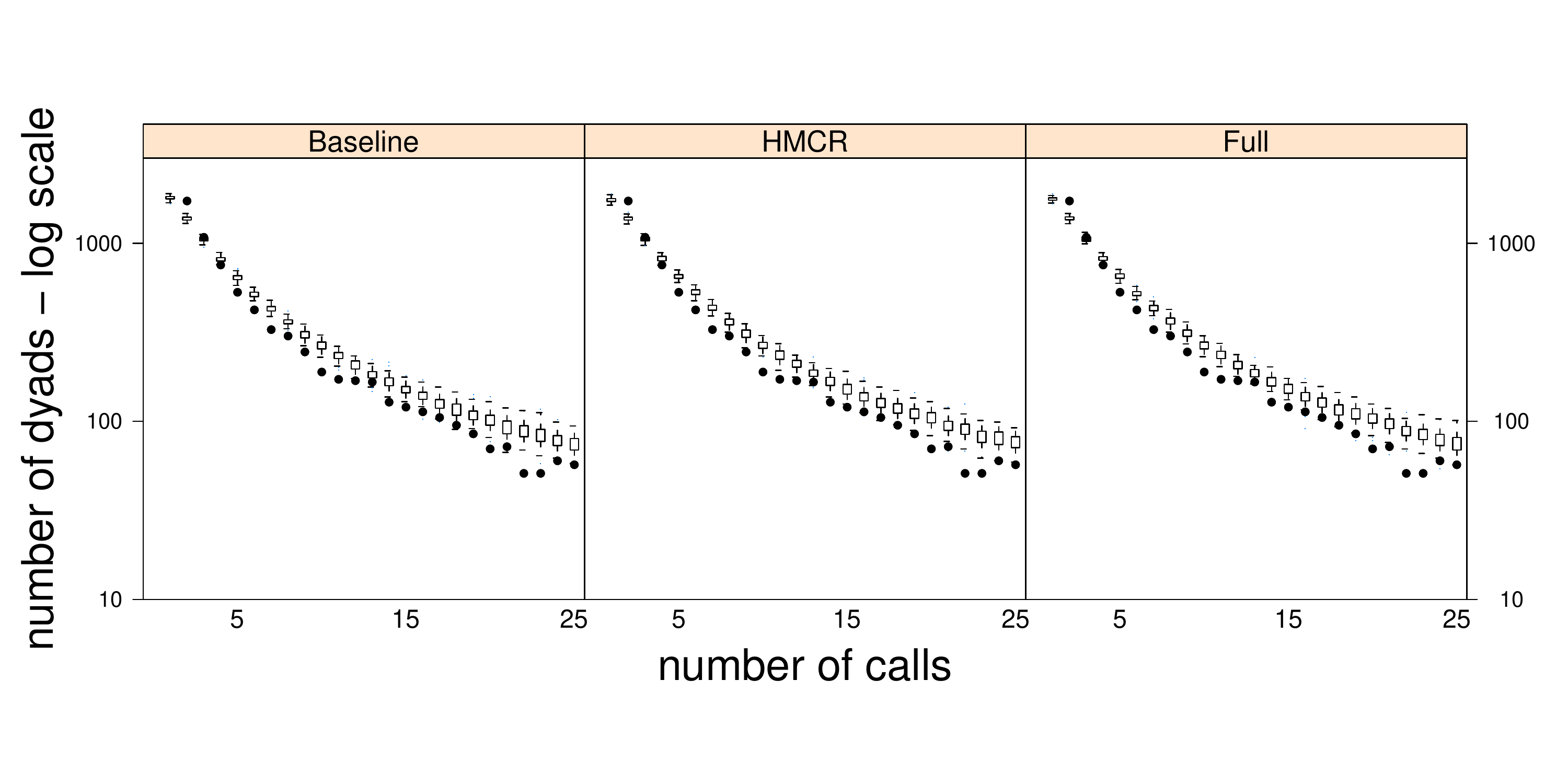}
}
\caption{Posterior predictive checks for holdout sample.  For each sub-figure, observed
  data are represented using dots, and the posterior predictive
distributions are represented by the  ``box-and-whisker'' symbols.}
\label{figure:PPC1}
\end{figure}

\begin{figure}
\centering
\includegraphics[height=96mm,width=120mm]{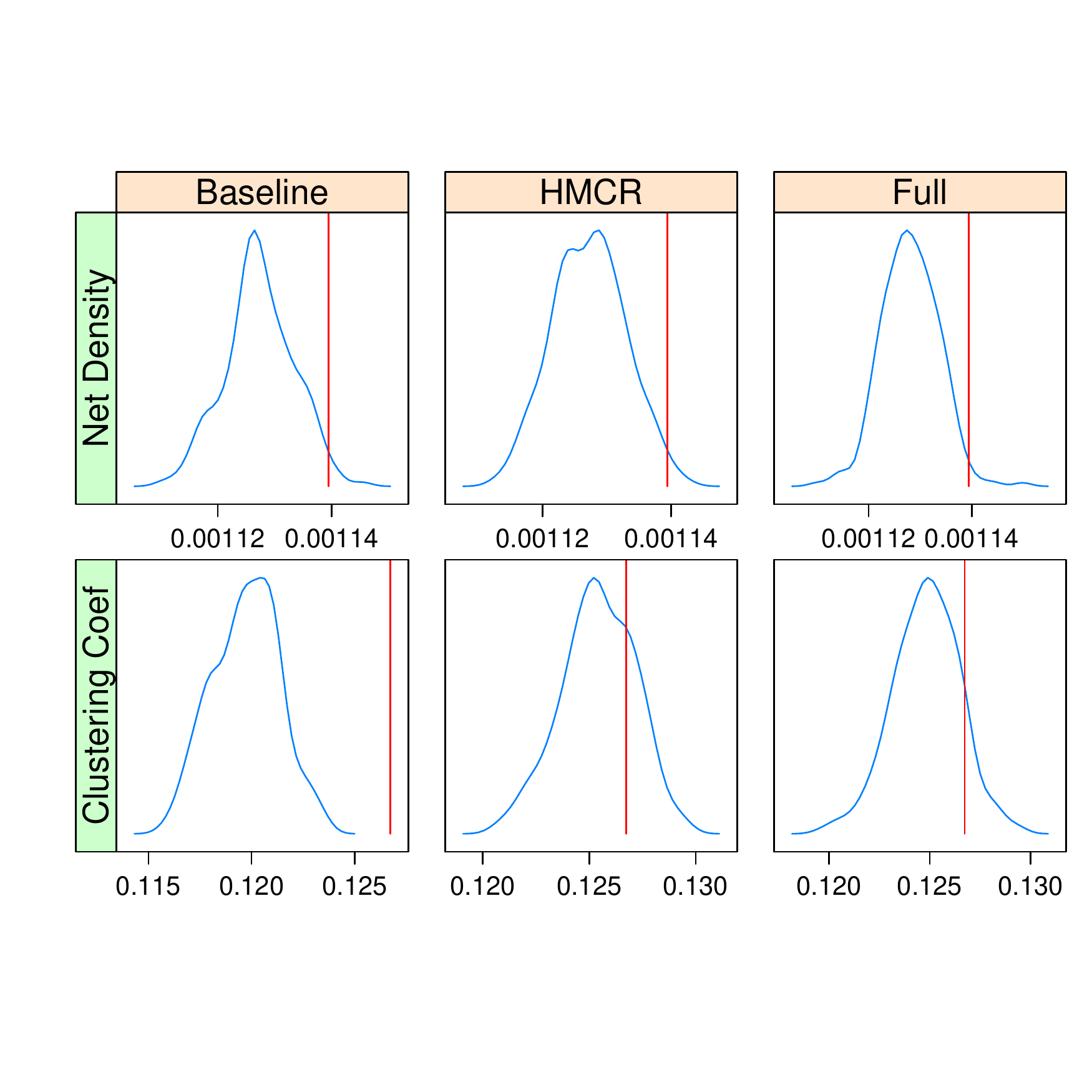}
\vspace{-0.8in}
\caption{PPCs for the density and clustering coefficients.}\label{figure:densClustPPC}
\end{figure}

Posterior predictive checks allow us to evaluate how well our model represents the data-generating process \citep{Rubin84,GelmanMengStern96}.
Three of our PPC test statistics are the same as those used by
\citet{HunterGoodreauHandcock2008} to assess goodness-of-fit for social networking data:
the degree distribution, the dyad-wise shared partner distribution, and
the distribution of geodesic distances.  We also examine the histogram of the number 
of calls made within non-empty dyads, the density of the network, and the clustering 
coefficient for the network.  All of our PPCs in
this article are with respect to the 26-week \emph{holdout} sample.
Figure \ref{figure:PPC1} shows the results for the distributional PPCs, and Figure \ref{figure:densClustPPC}
shows the PPCs for the density and clustering coefficients.  The
x-axis in each panel is the count of individuals or dyads, and the
y-axis is the log proportion of those individuals with each count.  The dark dots represent the
log probabilities generated from the actual dataset, and the
box-and-whisker plot represents the distribution of log probabilities
across the simulated datasets.  Figure \ref{figure:densClustPPC} shows the PPCs for the 
network density and clustering coefficients; the vertical line is the observed value.

At first glance, it might appear that all of the models replicate the
actual datasets rather well.  The reason that even the Baseline model
does as well as it does is that most of the value from posterior prediction
comes from inferring whether a dyad is open or closed.  Simply looking
at whether a dyad is empty or non-empty provides a lot of
information about the likelihood of future emptiness, because
non-empty dyads must be open.  However, closer examination reveals
that the Baseline model is not well calibrated at all.  The
``actual'' dots lie far outside the whiskers for the predictive
distributions for many of the counts.  The two models that involve some
kind of latent space structure fit better on these test
statistics.  However, we do not see much
difference between the HMCR and Full models.  This suggests that the
value of the latent space is more in predicting the potential existence of an interaction (whether the dyad is open or closed), than in predicting
the contact rate.

In addition to assessing model fit in aggregate, we also care about how well the model performs at the dyad level.  Our approach here is to predict which of the dyads that are empty during the calibration period become nonempty in the holdout period.  Empty dyads all have the same observed data pattern, so there is no obvious way to differentiate among them.   We can, however, use the latent space structure and a straightforward application of Bayes's Theorem to compute posterior distributions of unobserved parameters, and then use those probabilities to rank dyads in terms of those most likely to generate interactions during some future period of any duration we want.  To assess the predictive ability of a model, we first identify, individual by individual, the top $Q\%$ lift (or the top $q$ lift, where $q=Q/100$) most likely, previously uncontacted, individuals observed to contact during the holdout period\footnote{\baselineskip 12pt For any individual $i$, the top q lift requires rank ordering all potential customers $j \neq i$ based on the predicted probability of interaction.  For a holdout sample, the top $q$ lift of this rank ordered list is equal to the proportion of the top $q$ customers for whom we observe interactions with $i$, divided by the proportion of total interactions made by customer $i$.}.  
For a completely random or naive model, the percentage of  the top $Q\%$ most likely empty dyads to become non-empty should be equal to $q$.  For any other model, if the value for the top $Q\%$ metric is greater than $q$, then the model provides some "better-than-chance" predictive value.  The use of the $Q\%$ lift metric ensures that the maximum of this value is always equal to one.  

Table \ref{table:lift} presents these lift metrics for different models and values of $Q\%$, and different calibration/holdout samples.  Results are presented for all three model variants, with $D=2$  for the latent space models.  In addition, we present results for a ``Condition on Observed'' prediction rule, under which empty dyads are to remain empty in holdout, and non-empty dyads remain non-empty in holdout.  The ``Geodesic Distance'' model ranks dyads according to their geodesic distances, as in \citet{KossinetsWatts2006} (we break ties in two different ways:  randomly, or based on the total number of observed interactions along the path).   The lift metrics suggest that both the Baseline model and the ``Condition on Observed'' rule do exactly as well as one would expect from random selection.  This is because they both assumes that there is no network structure among individuals in the dataset, and thus all empty dyads are considered to be identical.  In contrast, in the two latent space models, some dyads are more likely to contact each other than others.  By sorting the empty dyads according to their posterior latent distances, we no longer assume that all empty dyads are the same.  Thus, we can improve on dyad-level prediction dramatically.  We do not, however, see any substantive differences between the HMCR and Full models, suggesting that, in this application, all of the action is on the open/closed probability and not on the contact rates.  Nevertheless, our results indicate that the use of the latent space structure for networked data is a better model than assuming independence across dyads.\footnote{There are of course many different ways to predict link formation (known in the machine learning community as ``link mining''), such as the Katz Score \citep{Katz1953} and the SimRank algorithm \citep{JehWidom2003}.  \citet{GetoorDiehl2005} provides a detailed review of link mining methods, and \citet{LibenNowellKleinberg2007} compare the performance of some of them.  We compared the predictive ability of our latent space approach against some of these methods, and found that while our model did best when tests were more discriminating (low $Q$), the other models ``caught up'' when $Q$ was increased.  However, our model offers behavioral intuition (see Section \ref{section:interpretation}) that machine learning algorithms cannot provide, and we are willing trade off some predictive power for managerial interpretability.  Nevertheless, our objective in predicting future link formation is only to demonstrate the value of accounting for latent network structure when modeling interactions data.}  

\begin{table}\centering
\begin{tabular}{lrrrr|rrr|rrr}
&&\multicolumn{9}{c}{Duration of calibration (holdout) period}\\

&&\multicolumn{3}{c|}{13 (39) weeks}&\multicolumn{3}{c|}{26 (26) weeks}&\multicolumn{3}{c}{39 (13) weeks}\\
&Q\%=&0.1&0.2&1.0&0.1&0.2&1.0&0.1&0.2&1.0\\
\hline
  Baseline&& .001 & .002  &.010& .001 & .002 &.010& .001 & .002 &.010\\
  HMCR && .100 & .112 &.141& .133  &.138 &.178& .153 & .154 &.177\\
Full && .100 &.109 &.146& .132 & .135 &.157& .154 & .159&.197\\
\hline
Condition on Observed && .001& .002&.010& .001& .002&.010& .001& .002 &.010\\   
Geodesic - random tiebreak && .050 & .088  &.166& .036 & .067 &.149& .025 & .051 &.198\\
Geodesic - \# calls tiebreak && .064 & .110 &.186& .056 & .098  &.190& .046 & .083 &.206\\
\end{tabular}
\caption{Percentage of the top $Q\%$ of the empty dyads (in calibration period) that are most likely to become nonempty during holdout period, that actually did become nonempty during the holdout period.   Reported values are posterior means; credible intervals are removed for space and clarity.}
\label{table:lift}
\end{table}

\section{Scalability and computation}
\label{section:scalability}

Having demonstrated the contribution of latent space models, we now turn to the issue of scalability and computation.  The amount of computational improvement one can expect from using DP priors on the latent space depends on how well we can cluster dyads into groups that have the same data and parameters.  In networks in which every dyad generates a different observed outcome (e.g., if the network is dense and the observed value is continuous), the likelihood for each dyad will have to be computed separately, and a discrete representation of the latent space will have little effect.  However, the density of many (if not most) social networks tends to be very low.  Even if non-empty dyads generate data on a continuous domain (as in our China Mobile example), there are so many empty dyads, all with the same data, that the number of distinct likelihoods to compute is much lower than the total number of dyads in the network.  If the observed data is discrete, then even more aggregation is possible.  Of course, aggregation according to observed data is standard practice when a model is homogeneous, or marginal likelihoods are available in closed form. The DP prior lets us group observations with similar \emph{latent} parameters as well.

The number of likelihood evaluations at each MCMC iteration depends on two factors:  i) the number of groups with distinct data patterns (which in turn depends on the size and density of the network); and ii) the number of mass points for each realization from the Dirichlet process.  Dyads with the same $z_i,z_j$  pair, and the same value of $y_{ij}$, must have the same likelihood, since they have the same data and same parameters.  As long as we keep track of the number of dyads with each $z_i,z_j$ pair, we can compute the log likelihood for that pair once for each $y$, and multiply by the number of dyads with that pair and that $y$.  Among all the data zeros, there are only $\binom{k}{2}+1$ possible likelihood values.  If $k$ is less than $N$, there is a computational saving, even if all of the non-zero values of $y$ are different (as happens when $y$ is continuous).  If $y$ is discrete (so $\mathcal{C}_y$ is the number of distinct values of $y$), there are at most  $\left(\binom{k}{2}+1\right)\mathcal{C}_y$ possible likelihoods.  For a continuous $y$, but with a large number of zeros, the number of possible likelihoods is $\binom{k}{2}+1$, plus the number of nonzero $y$'s.  Clearly, the more distinct observed data patterns there are, the less one can take advantage of the discretization of the latent space that is generated by the DP.  But in the social networking applications that are common in marketing, networks are often very sparse, so we have at least one very large group of dyads with the same data.

To assess just how much computational savings there is, consider the Full model in the telephone call example.  The calibration dataset
has $12,617$ non-empty dyads; likelihoods for each of these dyads must be computed individually.  The mean of $k$ is $530$, so there are $140,186$
distinct distances between mass points on the latent space.  Instead of computing $11,413,973$ separate likelihoods for each of the empty dyads, we only need to compute $140,186$ of them.  Thus, the number
of likelihoods to compute at each MCMC iteration is $152,803$.   This
represents a $98.7\%$ reduction in computational requirements.

The extent to which our method can scale for datasets with many more individuals (large $N$) depends on how both the network density and $k$ change as $N$ increases.  Ultimately, these are both empirical questions, the second of which we cannot know up front because not only is  $k$  unobserved, but it can be influenced by the choice of $H_0$ and $\alpha$.  However, the expected number of mass points can be asymptotically approximated as $E(k)\approx\alpha\log\left(\frac{\alpha +N}{\alpha}\right)$ \citep{Antoniak1974,Escobar1994}.  Thus, if $\alpha$ is small, the expected number of mass points is also expected to be small, but it will grow for larger datasets.  If $\alpha$ is large, the number of mass points for smaller datasets might be larger, but this number will not grow as quickly for larger datasets.  To test how well this approximation works in practice, we estimated the full model using successively larger subsets of our original network.  We then fixed $\alpha$ at three different values:  0.5, 20 and 300 (instead of placing a weakly informative prior on $\alpha$, as we did in the main analysis). We also computed the total number of likelihood computations for each sweep of the Gibbs sampler, which is just $\binom{k}{2}$, plus the number of non-empty dyads in the dataset.

\begin{figure}[tp]
\centering
\includegraphics[scale=0.9]{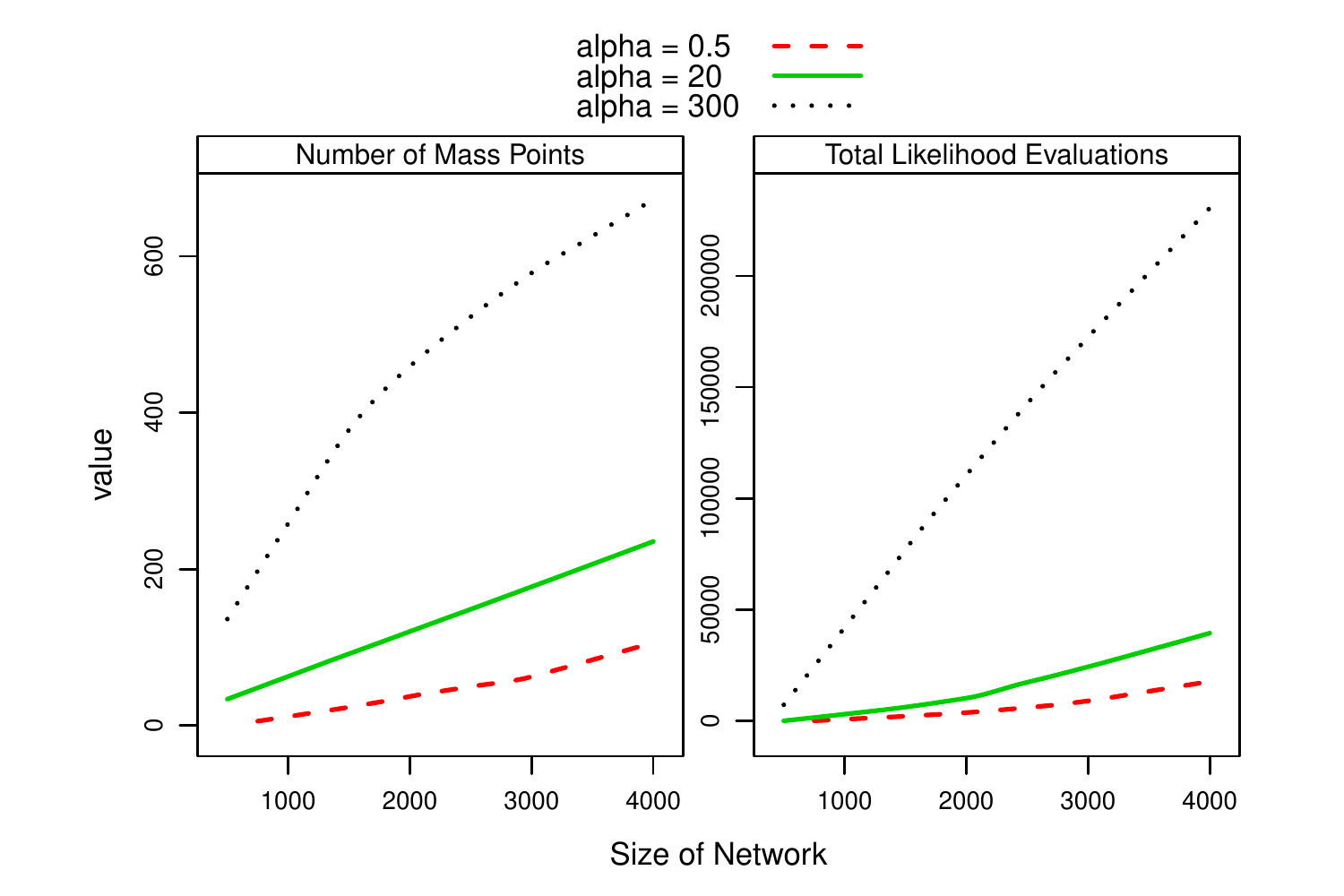}
\caption{Posterior means for number of mass points, and total likelihood computations, for subsampled networks.}\label{figure:scalePlots}
\end{figure}

Figure \ref{figure:scalePlots} plots the posterior mean of $k$ (the number of mass points) and the total number of likelihood evaluations, against the size of the network.  For the number of mass points, we observe the expected pattern.  For small $\alpha$, the number of mass points is small, but the incremental number of mass points grows with network size.  For large $\alpha$, the number of mass points is large, but the incremental change goes down with network size.   The asymptotic approximation suggests that incremental computational effort would decrease even more for even larger values of $N$, even though the number of total dyads continues to grow quadratically.   In terms of total computation, for low $\alpha$, the number of computations increases more rapidly with $N$ than for higher values of $\alpha$, but when $\alpha$ is high, the relationship becomes more linear.  Even though the number of dyads grows quadratically with $k$, larger networks will tend to have larger number of nonempty dyads. For low $\alpha$, computation grows faster than linear, but the number of latent dyads is low to begin with, because of the increased clustering.  Collectively, our results suggest that, if using a DP prior is not computationally feasible for a particular dataset, the incremental effort likely comes from the inability to aggregate the observed data, and not from an inability to aggregate the latent parameters.  Of course, this is no different from scalability problems faced by Bayesian hierarchical modelers who use MCMC to update model parameter estimates from non-networked data.

\section{Interpretation and usefulness of the latent space}\label{section:interpretation}
\begin{figure}[tb]
\centering
   \includegraphics[scale=0.57,angle=90]{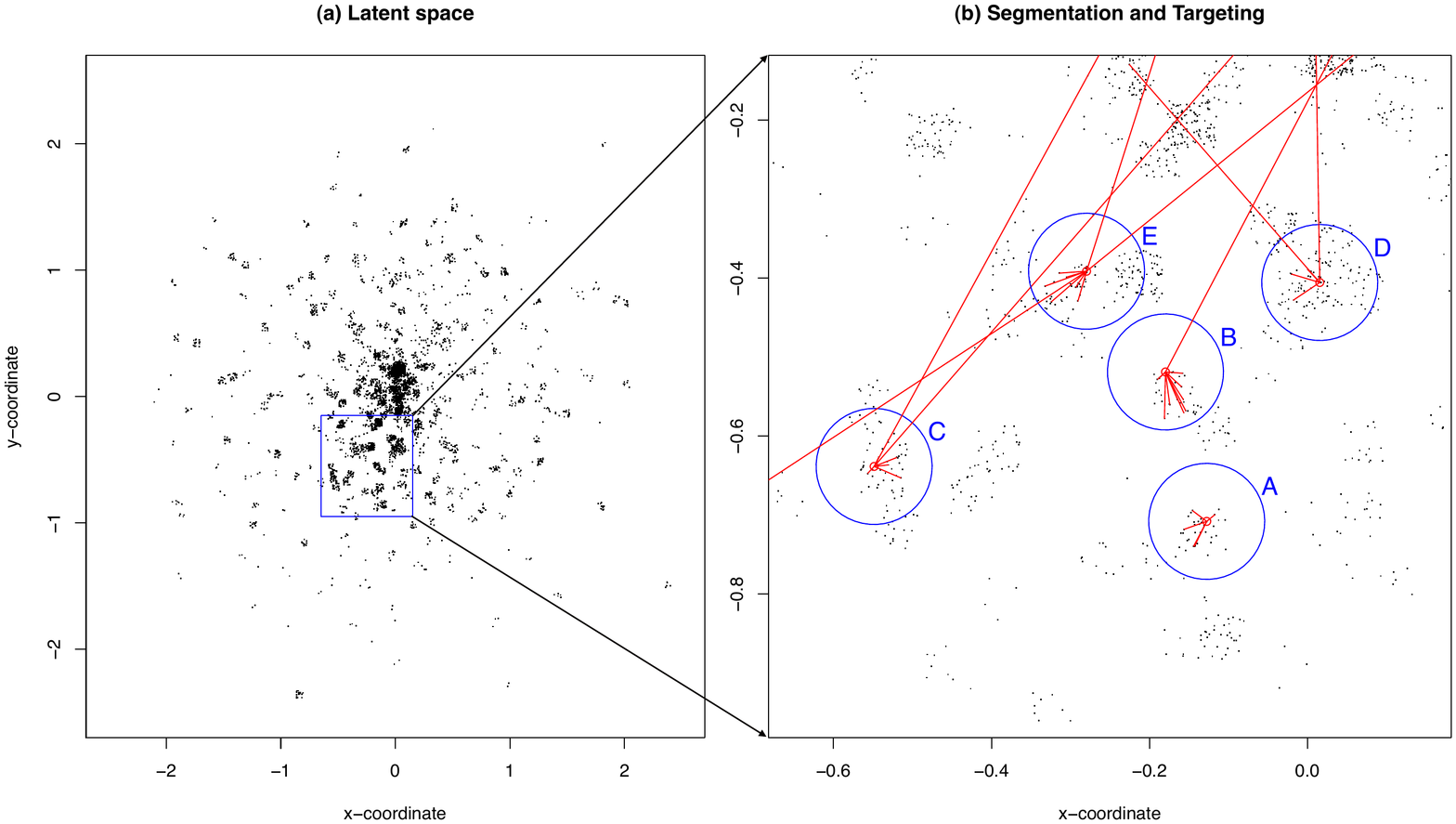}
\vspace{-0.75in}
\caption{Illustration of the full latent space, and connections among some selected customers.   In both panels the black dots represent  jittered locations for individual customer on $\{x,y\}$ coordinates.  Lines represent observed connections among customers.   In panel (b),  we labeled several customers who have hypothetically adopted a new product.   Circles around customers are used to identify other customers who may be similar to the targeted customers and therefore have similar adoption likelihoods.}\label{figure:latentspace}
\end{figure}

In Figure \ref{figure:latentspace}(a), we plot a single draw from the joint posterior distribution of the latent coordinates from the Full model with $D=2$ (we chose the draw with the largest conditional likelihood).   Each person in the network occupies a position in the latent space, and we define a cluster as all individuals who share the same coordinates in the latent space (this is akin to two observations having the same mass point in a realization from a mixture of Dirichlet processes, and we counted 593 such clusters in this realization).  But since multiple individuals are located at the same coordinates, to aid visualization we "jitter" the individual locations of customers by adding a small amount of random noise  (drawn from a uniform(-0.03,0.03) distribution) to each coordinate.  The scale labels on the axes are included to help reference certain parts of the space, and do not have a concrete interpretation themselves.  In Figure \ref{figure:latentspace}(a), we see that there is considerable clustering, with distinct ``super-clusters'' (clusters of clusters, or clusters closer to other clusters), of individuals on the latent space.  Also, there are some clusters that contain only a few customers, and which are quite separate from the rest of the network of customers, such as the one at $\{x,y\}$ coordinate (-0.9,-2.2).  Note that the latent space is a random variable, so this figure represents just one possible configuration of the individuals into clusters.  Figure\ref{figure:latentspace}(b) ``zooms in'' on a small partition of the latent space.  Having provided and discussed a graphical depiction of latent space, the next question becomes: what use is this to marketers?  We examine this in the context of segmentation and targeting.  

\paragraph*{Segmentation and targeting using interactions data} 

Segmentation and targeting is central to the development of effective marketing strategy.   The fundamental idea behind segmentation is to find people who are similar to one another, with the assumption that they will respond in similar ways, and therefore can be targeted using similar methods (e.g. the same price discount, the same promotion, or same advertising copy).   In our study we reveal two ways marketers can use network based data (our interactions observations) in practice.  As is well documented by authors such as \citet{NovakdeLeeuwMacEvoy1992} and  \citet{HillProvost2006}, following observed interactions to or from customers who have already adopted a product or service can help identify other potential customers.  Their results show improvements in response rates, compared with methods using observed, traditional segmentation and targeting bases.    While these network-based methods are powerful tools for eliciting new customers, our graphical representation of the latent space highlights that there are sometimes interactions among customers who are quite different from one another.   We see this in Figure \ref{figure:latentspace}(b). In this figure, we identify five individuals for whom a marketer might have some specific information (e.g., an existing customer, or a respondent to a promotion).  The lines radiating from these individuals represent \emph{observed} links in the dataset.  We also placed circles of common radius around these focal individuals.  

Network marketing tactics that ``follow the links," would use the lines to determine the next potential customers to target.  While this is useful in reaching new clusters, there are many marketing tactics that have nothing to do with following links or word of mouth, and instead depend more on understanding which customers can be grouped into more homogeneous segments.  Given the interpretation of the latent space as representing similarity, anyone in close proximity (within the circle) to a focal customer should also be a target.  Although most observed contacts also occur within a cluster,  there are certainly interactions among dissimilar individuals as well.  As an example, consider a marketer of trendy casual clothing, targeting a college student who interacts with two people:  a classmate at the same college, who shares similar demographic traits such as age, education, gender, values; and an older relative with whom there is a closer personal relationship, but nothing else in common in terms of purchase patterns.  While the college student might have identical observed interaction patterns with his classmate and with the relative, he shares many more common friends with his classmate than with his relative.   Our model places the student closer on the latent space to his classmate than to his relative, and the relative is closer to her own friends and others in her social circle.   This is useful for marketers to be aware of, because the classmate and the relative represent quite different marketing prospects.  If the marketer were to identify prospects based on the observed interactions alone, however, he could be targeting the relative, and her friends, who are unlikely to behave in the same way as the focal customer (the student).  
Targeting these prospects incurs additional costs with little expected return.  In addition, there are many individuals within the circle who never talk to our focal customer, but still ``travel in the same social circles,'' or who might otherwise be exposed to, or susceptible to, similar marketing activities.

We can compare the use of targeting based on proximity in latent space, with targeting based on geodesic distance\footnote{\baselineskip 12pt A third perspective is that perhaps one should first follow connections within the clusters, then the similar customers, and then follow the geodesic links outside of some cluster}.    In fact, many more customers can be identified for targeting than if one were to use a geodesic-type distance metric represented by observed interactions. We calculate that if one were to follow the first degree geodesic distance, on average the marketer would expect to reach eight customers (rounded up from 7.56) .   If this data were available and the marketer also included in the target set the "second degree," or friends of friends, the marketer then expects to reach on average 97 customers.  Drawing a circle of radius equal to 0.1 around the customer, the marketer may expect to reach, on average, 232 customers\footnote{\baselineskip 12pt Of course, this depends on the size of the circle, but since customers in the same cluster occupy identical coordinates, even a circle of minimal radius gives us the result that more customers are identified for targeting, on average, than the first-degree geodesic.}.  Since homophily implies that similar individuals are more likely to interact, then targeting based on latent space means that the customers identified for targeting are more likely to be similar to the focal customer.  The geodesic distance in some cases could be connections which span much of the space and therefore may lead to leads that are substantially different than the original customer.  For example, in Figure \ref{figure:latentspace}(b), the customer labelled "C" has seven interactions, but two of these interactions are to customers who are at substantially different locations in the latent space.  Given the desire for marketers to find customers similar to the focal customer, we assert that it is better to target those customers close to the labeled customers in the network.  The latent space model presents an opportunity to refine these targeting methods, and using the Dirichlet process to model the latent space makes the approach computationally feasible for marketing data. 

\paragraph*{Extracting information from limited data} 

Another advantage of using our probability modeling approach is that the \emph{interpretation} of the posterior latent space is only loosely dependent on the kind of data that one uses to estimate it.  Of course, larger, richer datasets, collected over longer periods of time, might lead to better posterior distributions, but the data itself could really be anything, as long as the underlying data-generating process is dyad-specific, and depends on similarities in a monotonic way (events are more likely if latent distances are small).   The data that is available could be limited in terms of time (a censored dataset), or by the fact that observed cell phone interactions are only one of many possible means of communication.   Interactions occur among customers at heterogeneous rates, and since it is not practical for marketers to wait extended amounts of time to see if interactions will occur among customers, it may be that some interactions that exist among customers occur at such a rate that they may not be observed in a small observation window.  One might be tempted to treat the addition or subtraction of observed interactions as evidence of nonstationarity.   The Bayesian approach to data analysis makes it straightforward to update our estimates of the latent space as new data becomes available, even when the space itself is stationary.  Therefore, what \citet{KossinetsWatts2006} refer to as  an "evolving social network" may, in effect, be an artifact of the censoring of the data.  All the observed data does is provide some clues from which the “true” underlying similarities must be inferred.   

Related to censoring, one of the concerns about using data like phone call records is that they do not necessarily represent the universe of interactions among the population.  Unless customers reveal all of the people with whom they ever interact with, observed data cannot be an authoritative document of the underlying social network.    There may be other modes of communication, such as email or face-to-face contact.  For example, \citet{AnsariKoenigsbergStahl2008} consider the case of a Swiss music sharing website, where the connection between users could be described alternatively as ``friendship,'' ``communication,'' or ``download.''  So what value does using only a network of cell phone calls have, if it is an incomplete representation of all interactions?  We can think of any ``true'' observed interaction network as a population of subnetworks, and each observed network (e.g., the cell phone data), as a single draw from that population  \citep{Gelman2007}.  We then treat that observed network as a single data point, and use it to update our beliefs about the structure of the latent space.  If we had observed another mode of communication first, we might get a different posterior latent space but, in any event, the posterior of the latent space after observing one network becomes the prior before observing the next.  

\paragraph{Focus on Diffusion and WOM}  We see two key contributions of the latent space useful to marketers managing the diffusion of innovation of information through customer networks.  The first involves the concept of "Word of Mouth" (e.g. \citealt{Arndt1967}).  While contagion could occur via non-explicit advocacy (e.g. fashion can be seen by people that one does not interact with), explicit communication is well regarded as an important source of information for customers.   WOM is at the heart of models of information diffusion in networks (e.g. \citealt{GoldenbergLibaiMuller2001}).  As testimony to the importance of consumer reviews, there are many services and organizations focused on collecting and presenting such information on just about every product or service.  Of key interest in the WOM literature is how the network structure affects diffusion patterns.    The contribution of our work is in considering that WOM may work via some geodesic distance versus distance measured in latent space.   From relations data, only the geodesic distance can be studied, but there is likely considerable value to considering distance in latent space as a channel for WOM.

The second area where the latent space model could be useful in practice is in identifying influential customers.   
The concept of a market maven, or opinion leader (\citealt{KatzLazarsfeld1955,FeickPrice1987,IyengarVandenBulteValente2008,KratzerLettl2009}) has frequently been studied in the context of diffusion research.   However,  recent research challenges the notion that such influentials are the primary reason for "global cascading influence" \citep{WattsDodds2007}, i.e. the contagious diffusion of innovation or information throughout an entire network.  These insights suggests that it is vital for marketers to understand how influence can occur among all customers, and that there is more to WOM marketing than focusing attention on just influentials.  Observations from practice certainly seem  to support this, from the popularity of services such as BzzAgent, and Procter \& Gamble's Vocalpoint, which are not selective about recruiting only "opinion leaders," but rather would prefer more people in their network.

\paragraph*{Further research opportunities} 

The sociological theory of homophily, coupled with the 
latent space framework, yields a stochastic representation of the relative latent characteristics underlying interactions data.  The latent characteristics are represented on a latent space, and we propose a Bayesian nonparametric for the latent space using Dirichlet processes.  Latent spaces are well known in social network analysis, and Dirichlet processes and probability models are known in marketing.  However, due to the computational obstacle involved in estimating larger networks, the concepts have not yet fully integrated across disciplines.  Our research lowers this obstacle, and makes probability modeling more accessible to marketing researchers who possess data on customer interactions.  This approach maintains the properties of interdependence, heterogeneity and interpretability, a goal that is harder to accomplish with extant classical or machine learning approaches.

We readily admit that our interpretation of the latent space, and our suggestions on how to use it, depend on an acceptance of two premises.  First, we need to believe that similarities drive interactions, and so one can use interactions to infer similarities.  We use the volumes of research of homophily to support this contention, but we have not tried to test this directly.  Second, our recommendation that managers consider using latent distance, rather than observed geodesic distance, to segment and target customers assumes that similar individuals have correlated purchase preferences or behavior.  This is a premise that could be tested, and we hope that both researchers and practitioners will undertake that challenge.  Unfortunately, the data that we have at our disposal does not allow us to follow that path, but we would like to describe briefly how we think this might work.

The output of the latent space model is a posterior distribution of configurations on a latent space.  Figure \ref{figure:latentspace} is one such configuration.  Although the distances among individuals in a single configuration do not have a physical interpretation, we can still treat them as distances in a statistical sense.  So we could draw upon the methods of hierarchical spatial modeling to infer a correlation structure among individuals, similar to those described in \citet{BanerjeeCarlinGelfand2004}.  An example of this kind of treating a non-physical distance as a physical one is \citet{YangAllenby2003}, who computed a demographic distance between people based on profiles of personal characteristics.  So just as they modeled correlations in preferences as functions of observed geographic and demographic distance, we propose modeling these correlations as functions of \emph{latent} distances.  We hypothesize that since observed interactions represent only one possible path for the sharing of information, it is latent distance, rather than geodesic, demographic or geographic distance, that would best predict these correlations.  In order to conduct a test like this, one would need two types of data for the same set of people:  dyad-level interactions data to infer the latent space, and individual-level purchase data to see if latent distance explains correlations in purchase behavior.  As more and more business is conducted through mobile communications devices, we anticipate that data like this will become more available.  A corollary to this research stream would be to incorporate individual-level demographics or covariates into the latent space model, and to better understand how that information might complement interactions data in understanding purchase behavior.

\appendix
\section*{Appendices}

\section{Choosing dimensionality of the latent space}\label{app:LML}

There are a number of approaches one could take to selecting $D$, and finding a general method for choosing among different specifications of Bayesian hierarchical models, especially those that incorporate
nonparametric priors, remains an area of active research among
statisticians.  We believe that because of the abstract nature of the latent space, there is no ``correct'' value of $D$ that one needs to infer from the data.  \citet{Hoff2005} notes that one should choose the smallest value of $D$ that offers a reasonable model fit, erring on the side of parsimony.  Adding more dimensions improves model flexibility, but can also lead to over-fitting.  As far as objective measures go, he suggests examining the  log marginal likelihoods (LML) (we use the holdout LML for the Full model), as well as the posterior predictive checks (PPC)  for test statistics that capture important characteristics of the data (we discuss PPCs in Section \ref{section:ppc}). 

In Table \ref{table:LML} we present relative estimated LML of the HMCR and Full models, for different values of $D$, from both the calibration and holdout datasets.  Our estimates were generated using cumulant approximations, and adjusting for the discrepancy between posterior and prior support, using the methods proposed in \citet{Lenk2009}.   Results are normalized such that the reported estimate for the HMCR model, with $D=2$ is 0, for each of the calibration and holdout datasets, and then scaled by 1,000 for readability.   

\begin{table}[h]\centering
\begin{tabular}{c|rr|rr}
&\multicolumn{2}{c}{calibration}&\multicolumn{2}{|c}{holdout}\\
D &        HMCR &        Full &      HMCR &         Full  \\
\hline
2 &  0   &  -40  & 0   &    -337      \\
3 &  221   &  -310  &  -378  &   -690      \\
4 &   -396  &  -1,609  &  -190  & -1,792        \\
5 &    -1,550 &  -1,130  &  -3,183  &  -1,560      \\
6 &    -4,729 &  -829  &  -5,069  &  -1,573       \\
\end{tabular}
\caption
{Estimates of log marginal likelihoods (LML) of models, by dimensionality of latent
  space and model variant. Estimates are normalized with respect to the $D=2$ HCMR model for each dataset.}%
\label{table:LML}%
\end{table}%

In three of the four cases, $D=2$ is preferred, and in the remaining one, $D=3$ is preferred.  Also, we found essentially no difference among values of $D$ in posterior predictive checks.  So, following Hoff's advice, we use the $D=2$ models for subsequent analysis.

\section{Model specification for China Mobile example\label{app:posterior}}

In this appendix, we derive the data likelihood and hyperprior specifications for our Chongqing Mobile application.   Let $p_{ij}$ be
the probability that a dyad between $i$ and $j$ is open, and let
$\lambda_{ij}$ be the rate of contacts between $i$ and $j$ (assuming
exponentially-distributed intercontact times), if the dyad is open.
We also define an auxilliary latent Bernoulli variable $s_{ij}$ that
indicates whether a dyad is open ($s_{ij}=1$) or closed
($s_{ij}=0$).  To remain consistent with our general model
specification in the text, we use $y_{ij}$ to denote the vector of
observed intercontact times, and $y^*_{ij}$ to denote the count of
observed contacts.  Also, let $T$ be the duration of the observation period.

The data likelihood is similar to an ``exponential never triers'' model \citep{FaderHardieZeithammer2003}.  There are two ways a dyad could be empty (i.e., $y^*=0$):  the dyad could be closed ($s_{ij}=0$), or it could be open, but the contact rate $\lambda_{ij}$ is sufficiently low that there just happened to be no contacts during the observation period.  If we observe any contacts at all, we know the dyad must be open.  Therefore, the data likelihood is
\begin{equation}
f\left( y_{ij}| \lambda_{ij}, p_{ij}\right) = \left( 1-p_{ij}\right)\mathbf{I}\left[ y_{ij}^*=0\right] + p_{ij}\lambda_{ij}^{y_{ij}^*}\exp\left( -\lambda_{ij} T\right)
\end{equation}
We incorporate dyad-wise unobserved heterogeneity in $\lambda_{ij}$ by
using a gamma distribution with dyad-specific shape parameter $r_{ij}$
and dyad-specific scale parameter $a_{ij}$
\begin{equation}\label{eq:gammaPrior}
  f\left( \lambda_{ij}|r_{ij},a_{ij}\right) = \frac{a_{ij}^{r_{ij}}}{\Gamma\left(
      r_{ij}\right)}\lambda_{ij}^{r_{ij}-1}\exp\left( -a_{ij}\lambda_{ij}\right)
\end{equation}
After integrating over $\lambda_{ij}$,
\begin{align}\label{eq:margLike}
 f\left( y_{ij} | p_{ij}, r_{ij}, a_{ij}\right) = \left( 1-p_{ij}\right)\mathbf{I}\left[ y_{ij}^*=0\right] + p_{ij}\frac{\Gamma\left(
      r_{ij}+y_{ij}^*\right)}{\Gamma\left( r_{ij}\right)}
\left(\frac{a_{ij}}{a_{ij}+T}\right)^{r_{ij}}\left(\frac{1}{a_{ij}+T}\right)^{y_{ij}^*}
\end{align}
We will also use the reparameterizations $r_{ij}=\mu_{ij}^2/v$ and $a_{ij}=\mu_{ij}/v$, where $\mu_{ij}$ and $v$ are the dyad-specific mean and
common variance of the gamma distribution, respectively.  Linking back to our general model formulation in Section \ref{subsection:modelIntuition}, $\phi_{ij}=\left[p_{ij}, \mu_{ij}, v\right]$.  The definitions of these parameters are described in Section \ref{subsection:modelSpecifics}.

\subsection*{Hyperprior specifics}
For the choice of $H_0$, we decompose each latent coordinate into two components:
the distance from the origin (a ``radius'') and the location on the surface of a hypersphere
that has that radius. We then choose a $H_{0}$ that factors into a
prior on these two components.  In other words, we think of the
elements of $z_i$ in terms of their spherical, rather
than Cartesian, coordinates.  If the
Cartesian coordinates of $z_i$ (herein suppressing the $i$ subscript) are $z_{1},z_{2},...,z_{D}\,$, then its
polar coordinates are $\left(  \vartheta_{1},...,\vartheta_{D-1},\rho\right)
$, where $\rho$ is a distance from the origin and the $\vartheta^{\prime}$s are
angles, expressed in radians, such that $0<\vartheta_{1}<2\pi$, and $0<\vartheta
_{j}<\pi$ for $2\leq j\leq D-1$. We can then factor $H_0$ as
\begin{equation}
g_{0}\left(  z\right) =f\left(  \vartheta_{1},...,\vartheta_{D-1},\rho\right)  =f\left(
\vartheta_{1},...,\vartheta_{D-1}|\rho\right)  f\left(  \rho\right)
\end{equation}
Conditioning on $\rho$, we want to place a distribution on $\vartheta=\left(
\vartheta_{1},...,\vartheta_{D-1}\right)  $, such that there is a uniform
probability of being at any location on a $D-$dimensional hypersphere
with radius $\rho$. This is achieved by letting $f\left(
  \vartheta|\rho\right)  $ be a multivariate power sine distribution
\citep{Johnson1987,NachtsheimJohnson1988}, where
\begin{equation}
f\left(  \vartheta\right)  \propto\prod_{j=1}^{D-1}\sin^{j-1}\vartheta_{j}%
\end{equation}
Thus, $\vartheta_{1}$ has a uniform distribution, $f\left(  \vartheta_{2}\right)
\propto\sin\left(  \vartheta_{2}\right)  ,$ $f\left(  \vartheta_{3}\right)
\propto\sin^{2}\left(  \vartheta_{3}\right)  $, and so forth. \citet[ch~7]{Johnson1987}
proposes some algorithms for simulating from a multivariate power sine
distribution.

For $f\left( \rho \right)$, recall that $\rho$ is defined on
the positive real line, with $E\left(  \rho\right)  =1$. We also need the ability to
trade off tail weight (probability of draws of $z$ being far away
from the origin) against kurtosis (likelihood of draws of $z$
being clustered around the origin).  Beginning with the generalized Laplace
distribution \citep[sec. 4.4.2]{KotzKozubowski2001}, we center, and
then fold, at zero, to get
\begin{align}
f\left(  \rho\right) & =\left[  \kappa^{\frac{1}{\kappa}}\sigma\Gamma\left(
1+\frac{1}{\kappa}\right)  \right]  ^{-1}\exp\left[  -\left(  \frac{\rho
}{\kappa \sigma}\right)  ^{\kappa}\right]  \text{, }\rho>0
\end{align}
where
\begin{align}
E\left(  \rho\right) & =\kappa^{\frac{1}{\kappa}}\sigma\frac{\Gamma\left(  \frac
{2}{\kappa}\right)  }{\Gamma\left(  \frac{1}{\kappa}\right)  }
\end{align}
Setting $E\left(  \rho\right)  =1,$
\begin{equation}
\sigma=\frac{\Gamma\left(  \frac{1}{\kappa}\right)  }{\Gamma\left(  \frac{2}
{\kappa}\right)  \kappa^{\frac{1}{\kappa}}}%
\end{equation}
so the density of $\rho$, constrained so $E\left( \rho\right)  =1$, is
\begin{equation}
f\left(  \rho|\kappa\right)  =\frac{\kappa\Gamma\left(  \frac{2}{\kappa
}\right)  }{\Gamma\left(  \frac{1}{\kappa}\right)  ^{2}}\exp\left[  -\left(
\frac{\rho\Gamma\left(  \frac{2}{\kappa}\right)  }{\Gamma\left(  \frac
{1}{\kappa}\right)  }\right)  ^{\kappa}\right] \label{radii}
\end{equation}
The parameter $\kappa$ controls the tradeoff between tail weight and
kurtosis. If $\kappa=1$, $f\left(  \rho|\kappa\right)  $ reduces to an exponential
distribution, and if $\kappa=2$, $f\left(  \rho|\kappa\right)  $ is a
half-normal distribution. As $\kappa$ becomes large, the mode of $\rho$
becomes less and less peaked, and $f\left(  \rho|\kappa\right)  $ converges to
a uniform$\left(  0,2\right)  $ distribution. The ``correct'' value for
$\kappa$ is inferred through the estimation process, letting the data drive
the tradeoff between placing a mode on $\rho$ and bounding the
locations such that $\rho\leq 2$.  Note that this prior using the
multivariate power sine distribution and our restricted half-Laplace
distribution adds only one additional parameter to the model, compared
to an independent multivariate normal hyperprior, which adds many more.
\par
It turns out that $f\left(  \rho|\kappa\right)  $ is a special case of a power
gamma distribution. To see this, perform a change of variables so
$\varpi=\rho^{\kappa}$, $\rho=\varpi^{\frac{1}{k}}$ and $d\rho=\frac{1}{\kappa}%
\varpi^{\frac{1}{\kappa}-1}$. Then,
\begin{equation}
f\left( \varpi|\kappa \right) =\frac{\Gamma \left( \frac{2}{\kappa }%
\right) }{\Gamma \left( \frac{1}{\kappa }\right) }\varpi^{\frac{1}{\kappa }%
-1}\exp \left[ -\left( \frac{\Gamma \left( \frac{2}{\kappa }\right) }{\Gamma%
\left( \frac{1}{\kappa }\right) }\right)^{\kappa }\varpi\right]
\end{equation}
which is a gamma distribution with shape parameter $\frac{1}{\kappa}$ and
rate parameter $\left(  \frac{\Gamma\left(  \frac{2}{\kappa}\right)  }%
{\Gamma\left(  \frac{1}{\kappa}\right)  }\right)  ^{\kappa}$. This result
makes it easy to simulate values of $\rho$; just draw $\varpi$ from this gamma
distribution, and transform $\rho=\varpi^{\frac{1}{\kappa}}$. After simulating
values of $\vartheta$ and $\rho$, it is often convenient to convert $z_{ij}$ back to
its Cartesian coordinates. The elements of  $z$ can be
expressed as:
\begin{align}
z_{1}  & =\rho\cos\vartheta_{1}&
z_{j}  & =\rho\cos\vartheta_{j}\prod_{l=1}^{j-1}\sin\vartheta_{l},\text{ for }2\leq j<D&
z_{D}  & =\rho\prod_{l=1}^{D-1}\sin\vartheta_{l}%
\end{align}
\citep[ch~7]{Johnson1987}.

We selected the other hyperpriors to balance weak information content against numerical stability. Following \citet{EscobarWest95}, we place a weakly informative gamma prior on $\alpha$, with mean of 4 and variance of 80.  Since  $\beta \text{ and } v$ are all population-level
parameters that appear only in the definitions of $\phi_{ij}$, we can combine them all into a single parameter vector $\Xi$
(log-transforming parameters when necessary), with a multivariate
normal prior $\Xi_0$, centered at the origin, with covariance matrix $A=10I$.  Note that if $\kappa=2$, then $H_0$ is a multivariate normal distribution.  Since we were concerned about a mode of $H_0$ introducing too much prior information, we used a gamma prior with a mean of 3 and a variance of about 5.  Experimenting with alternative values led to no substantive effect.

\section{Estimation algorithm\label{app:SamplerAppendix}}

In this section we present the complete MCMC sampling algorithm for
the general latent space model.   The
parameters to be estimated are $\alpha$, $\kappa$, $\beta$, and $z_{i}$,
$i=1...N$. Recall that since $H$ is discrete, at each iteration there are only
$k$ possible values that any $z_{i}$ can take.

\subsection{Simulate $\alpha|\cdot$}

Let $r_{\alpha}$ and $a_{\alpha}$ be the parameters of the gamma hyperprior on $\alpha.$ Using
the algorithm proposed by
\citet{EscobarWest95}%
, do the following.
\begin{enumerate}
\item Starting with the current value of $\alpha$, draw a temporary variable
$\eta$ from a $\text{Beta}\left(  \alpha+1,N\right)  $ distribution.

\item Draw $\tau$ from a Bernoulli trial with probability $\frac{r_{\alpha}+k-1}{N\left(
a-\log\left(  \eta\right)  \right)  +r_{\alpha}+k-1}$

\item If $\tau=0$, draw $\alpha$ from a $\text{gamma}\left(  r_{\alpha}+k-1,a_{\alpha}-\log\left(
\eta\right)  \right)  $ distribution. If $\tau=1$, draw $\alpha$ from a
$\text{gamma}\left(  r_{\alpha}+1,a_{\alpha}-\log\left(  \eta\right)  \right)  $ distribution.
\end{enumerate}

\subsection{Simulate $\beta, v|\cdot$}

To simplify notation, we combine $\beta \text{ and } v$ into a single parameter vector $\Xi$.  The
conditional posterior distribution of $\Xi$ depends on the data
likelihood and the prior.  The data likelihood we care about here is the marginal likelihood in Section \ref{subsection:modelFormal}, after integrating over $\theta_i$, multiplied across all dyads (using our assumption
 of \emph{conditional} independence across dyads).  Note that $\phi_{ij}$ is a function of $\Xi, z_i \text{ and } z_j$.  The prior on $\Xi$ is a multivariate normal with mean $\Xi_0$ and covariance $A$. Thus, the log
conditional posterior (without normalizing constant) for $\Xi$ is

\begin{equation}
\log f\left(  \Xi|\cdot\right)  =\sum_{i=1}^N \sum_{j=i+1}^N \log f(y_{ij}|\Xi, z_i, z_j) -\frac{1}{2}\left(
\Xi-\Xi_0\right)  ^{\prime}A^{-1}\left(  \Xi-\Xi_0\right)
\end{equation}

We simulate $\Xi$ using a random walk Metropolis sampler \citep[ch. 3]{RossiAllenbyMcCulloch2005}.
\subsection{Simulate $\kappa|\cdot$}

We place a $\text{gamma }\left(r_{\kappa},a_\kappa\right)$ prior on
$\kappa$.  Let $\rho_{1:k}$ be the radii of the $k$ distinct values of $z$. Combining the
likelihood of the radii in \ref{radii} with the prior,
the conditional posterior distribution for $\kappa$ is
\begin{equation}
f\left( \kappa|\cdot\right)  \propto\left[  \frac{\kappa\Gamma\left(  \frac
{2}{\kappa}\right)  }{\Gamma\left(  \frac{1}{\kappa}\right)  ^{2}}\right]
^{k}\exp\left[  -\sum_{j=1}^{k}\left(  \frac{\rho_{j}\Gamma\left(  \frac
{2}{\kappa}\right)  }{\Gamma\left(  \frac{1}{\kappa}\right)  }\right)
^{\kappa}-a_{\kappa}\kappa\right]  \kappa^{r_{\kappa}-1}%
\end{equation}
There are many different ways to simulate from this univariate density. We chose to use sampling-importance
resampling (SIR) \citep{SmithGelfand1992}, but one might choose Metropolis, grid-based inverse CDF, or slice sampling
methods instead.

\subsection{Simulate $z|\cdot$}

This step, in which we draw each of the $z_i$ vectors from the mixture of Dirichlet processes (MDP), is an adaptation Algorithm 8 in
\citet{Neal2000}. We direct the reader there for an explanation of how and why the algorithm
works, but we present a summary here, using our terminology
and notation. The distribution of the $z_i$'s is discrete, so at each iteration of the estimation algorithm, there are only $k$ possible values that $z_i$ can take.  
Let $z^{\ast}=\left(  z_{1}^{\ast}\mathellipsis z_{k}^{\ast}\right)  $ define these $k$ distinct
latent coordinates, let $z_{-i}^{\ast}$ be the distinct
mass points when not including person $i$, and let $k_{-i}$ be the number of
distinct mass points in $z_{-i}^{\ast}$ when not including person $i$ ($z^{\ast}$ and
$z_{-i}^{\ast}$, and $k$ and $k_{-i}$,will differ only if $i$ is a ``singleton'' who is the only person located at $z_i$). At the current
state of the sampler, each person is ``assigned'' to one of the $z_{j}^{\ast}$,
in the sense that there is exactly one $j$ for which $z_{i}=z_{j}^{\ast}$. Let
$N_{j}$ be the number of people assigned to $z_{j}^{\ast}$, and let $N_{-i,j}$
be the number of people assigned to $z_{j}^{\ast}$, when not counting person $i$.

The algorithm involves choosing some number of proposal values (determined by a prespecified control parameter $m$) for each $z_{i}$. Define $f_i\left(  y_{i}|z_{i},z_{-i}\right)  $ as the likelihood contribution for all dyads that
involve person $i$, given the current values of $z_i$ assigned to $i$, and of $z_{-i}$ assigned to everyone
else. There are two cases that we need to consider. The first is if there is
some other person $i^{\prime}$ for which $z_{i}=z_{i^{\prime}}$ (i.e., $i$ is
not a singleton). In this case, draw $m$ proposal draws from $H_{0}$, call them $z_{k+1}^{\ast},...,z_{k+m}^{\ast}$, and let $\tilde
{z}=\left(  z_{1}^{\ast},...,z_{k}^{\ast},z_{k+1}^{\ast},...,z_{k+m}^{\ast
}\right)  $.  Intuitively, $\tilde{z}$ is the union of the set of all latent vectors that are already assigned to someone in the population, with the set of $m$ new proposal vectors.
Next, compute $f_i\left(  y_{i}|\tilde{z}_{j},z_{-i}\right)  $, for all
$j=1\mathellipsis (k+m)$. These are the likelihood contributions for all dyads involving
$i$, if $z_{i}$ were set to each of the values in $\tilde{z}$.  These ``proposal likelihoods'' form a set of weights that we use to draw a new $z_i$ for each $i$.  Thus, draw a new value for $z_{i}$ from $\tilde{z}$ using the following
probabilities:
\begin{align*}
\Pr\left(  z_{i}=\tilde{z}_{j}\right)  =
\genfrac{\{}{.}{0pt}{}{\omega\frac{n_{-i,j}}{N-1+\alpha}F_i\left(  y_{i}%
|\tilde{z}_{j},z_{-i}\right)  \text{ for }1\leq j\leq k}{\omega\frac{\left(
\alpha/m\right)  }{N-1+\alpha}F_i\left(  y_{i}|\tilde{z}_{j},z_{-i}\right)
\text{ for }k+1\leq j\leq k+m}%
\end{align*}
where $\omega$ is a normalizing constant (and does not need to be known for the purposes of random sampling). Thus, $z_{i}$ can take on the
value of any of the existing elements of $z^{\ast}$, or one of the $m$ new
candidate values. Which value is selected depends on three values: the
likelihood of the data for each $\tilde{z}_{j}$ (values of $\tilde{z}_{j}$
that yield a high likelihood are more likely to be chosen), the number of
other people who also are assigned to $\tilde{z}_{j}$ (coordinates where the
prior distribution has more mass are more likely to be chosen), and the DP
control parameter $\alpha$, which governs how close the DP prior on $z$ is to
$H_{0}$. If $i$ is a singleton, then there are only $k_{-i}=k-1$ elements in
$z^{\ast}$. In this case, draw $m+1$ candidate values from $H_{0}$, re-index them
so $\tilde{z}=\left(  \left\{  z_{-i}^{\ast}\right\}  ,z_{k}^{\ast
},...,z_{k+m}^{\ast}\right)  $, and select according to the probabilities
\begin{align*}
\Pr\left(  z_{i}=\tilde{z}_{j}\right)  =
\genfrac{\{}{.}{0pt}{}{\omega\frac{N_{-i,j}}{N-1+\alpha}F_i\left(  y_{i}
|\tilde{z}_{j},z_{-i}\right)  \text{ for }1\leq j\leq k-1}{\omega\frac{\left(
\alpha/m\right)  }{N-1+\alpha}F_i\left(  y_{i}|\tilde{z}_{j},z_{-i}\right)
\text{ for }k\leq j\leq k+m}%
\end{align*}

\singlespacing
\bibliography{../refs/braun_refs}

\end{document}